\documentclass[lettersize,journal]{IEEEtran}
\usepackage[T1]{fontenc}
\usepackage{newunicodechar}
\newunicodechar{‑}{-}
\newunicodechar{–}{--}
\newunicodechar{—}{---}
\newunicodechar{’}{'}
\newunicodechar{‘}{`}
\newunicodechar{“}{``}
\newunicodechar{”}{''}
\newunicodechar{→}{$\rightarrow$}
\usepackage{amsmath,amsfonts}
\usepackage{algorithmic}
\usepackage{algorithm}
\usepackage{array}
\usepackage{textcomp}
\usepackage{stfloats}
\usepackage{url}
\usepackage{verbatim}
\usepackage{enumitem}

\usepackage{multirow}
\usepackage{graphicx}
\usepackage{cite}
\usepackage{color}
\usepackage{soul}
\usepackage{caption}
\usepackage{subcaption}
\usepackage{lipsum}
\begin{document}

\setlength{\textfloatsep}{3pt}
\setlength{\intextsep}{3pt}
\setlength{\floatsep}{3pt}

\title{Tutorial on Closed-Loop and Application-Aware Digital Twins for Future Wireless Networks}

\author{
		\IEEEauthorblockN{
        Yuchen Liu,~\IEEEmembership{Member,~IEEE,} 
		Sara Garcia Sanchez,~\IEEEmembership{Member,~IEEE,}
		Mingzhe Chen,~\IEEEmembership{Member,~IEEE,} \\
            Xingqin Lin,~\IEEEmembership{Senior Member,~IEEE,}
            Berk Canberk,~\IEEEmembership{Senior Member,~IEEE,}
            Tommaso Melodia,~\IEEEmembership{Fellow,~IEEE} \\
 }

  \thanks{{
Y. Liu is with
} the Department of Computer Science, North Carolina State University, Raleigh, NC, 27695, USA (Email: yuchen.liu@ncsu.edu).}  
  
  \thanks{S. Sanchez {
is with IBM
}, NY, 10504, USA (Email: \protect\url{ Sara.Garcia.Sanchez@ibm.com)}.} 
  
\thanks{M. Chen is with the Department of Electrical and Computer Engineering and Frost Institute for Data Science and Computing, University of Miami, Coral Gables, FL, 33146, USA (Email: \protect\url{ mingzhe.chen@miami.edu)}.} 

\thanks{X. Lin is with the NVIDIA, Santa Clara, CA, 95051, USA (Email: \protect\url{ xingqinl@nvidia.com)}.} 

\thanks{B. Canberk is with the School of Computing, Engineering and the Built Environment, Edinburgh Napier University, Edinburgh, U.K. (Email: \protect\url{ b.canberk@napier.ac.uk)}.}

\thanks{T. Melodia is with the Department of Electrical and Computer Engineering, Northeastern University, Boston, MA, USA (Email: \protect\url{ t.melodia@northeastern.edu)}.}

}

\maketitle

\begin{abstract}
To support emerging applications such as extended reality, autonomous systems, and large-scale cyber-physical infrastructures, next-generation wireless networks must evolve beyond traditional communication paradigms toward tightly integrated and closed-loop systems. Digital Twin Networks (DTNs) have recently emerged as a key enabler of this transformation by bridging the physical and digital worlds through real-time, bidirectional interactions and adaptive control mechanisms. In this tutorial, we provide a comprehensive overview of DTN design and optimization across multiple system dimensions.
Specifically, we first introduce the fundamental design principles of DTN creation, including high-fidelity twin modeling, telemetry integration, and cyber-physical synchronization. We then present resource optimization and semantic-driven DTNs, where application-aware objectives and semantic abstractions guide efficient data processing, communication, and decision-making. Subsequently, we explore physical-layer optimization techniques for DTNs, emphasizing joint sensing, communication, and adaptive wireless resource control. Furthermore, we examine DTN-guided network management and feedback optimization, focusing on closed-loop learning, real-time control, and reinforcement-driven adaptation between digital twins and physical networks. Finally, we review ongoing standardization efforts and enabling tools that support the practical deployment of DTNs in real-world systems. This tutorial aims to provide a unified perspective on DTNs as an emerging paradigm for application-aware, AI-native, and closed-loop wireless systems, and to highlight key challenges and future research directions in this rapidly evolving field.

\end{abstract}


\section{Introduction}
\label{intro}

\begin{figure}[t]
    \centering
    \includegraphics[scale=0.142]{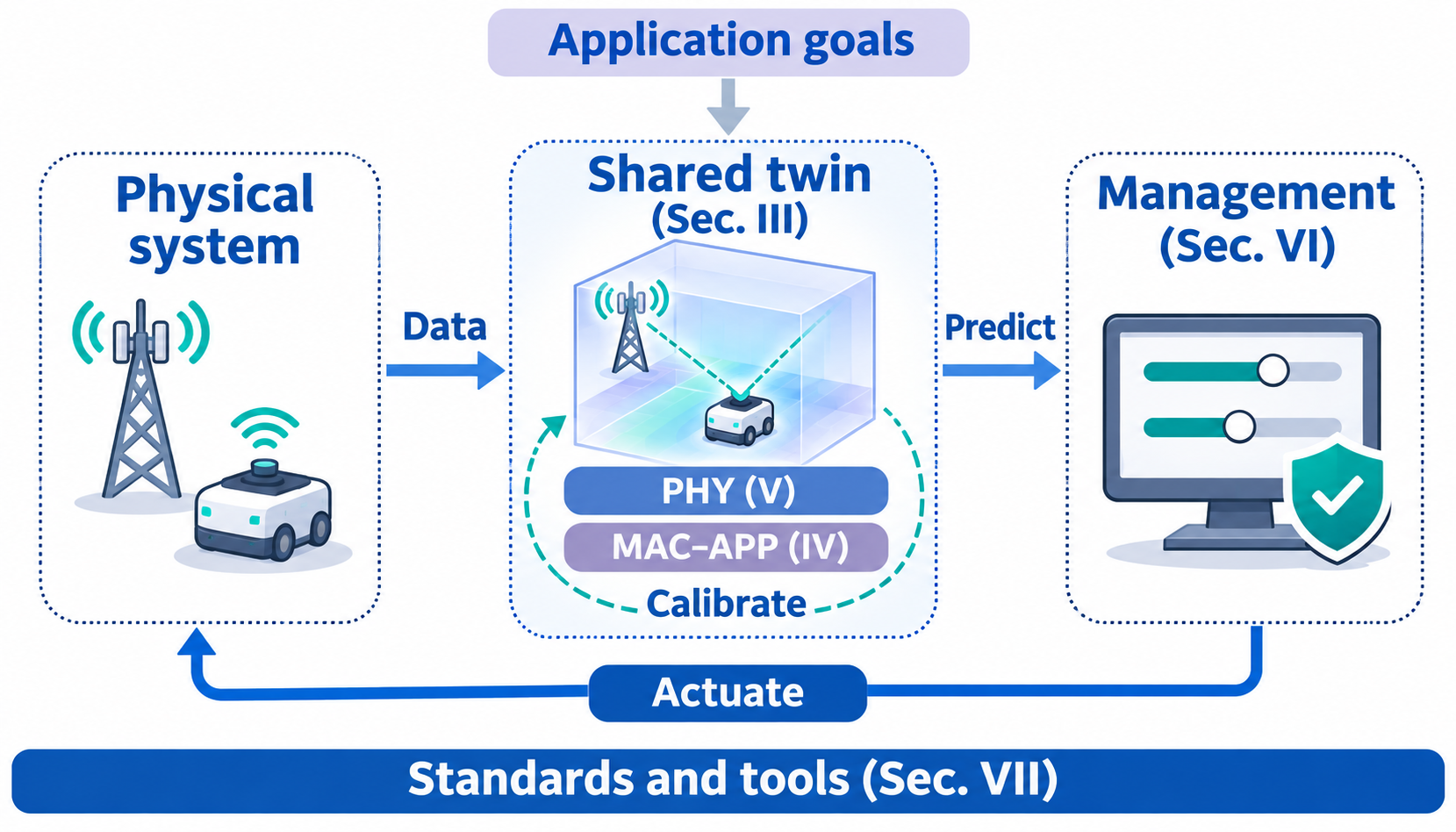}
    \caption{Tutorial overview: application goals guide shared twin models and cross-layer management through measurement, prediction, calibration, and actuation.}
    \label{fig:intro}
\end{figure}

Digital Twin (DT) technology has evolved from its origins in manufacturing and industrial automation into a foundational technology for cyber-physical systems. By maintaining a synchronized virtual representation of a physical entity, DTs enable monitoring, prediction, and intelligent decision making throughout the system lifecycle. Motivated by these capabilities, DTs have recently attracted significant attention in wireless communications and networking~{\cite{11217298}}, where they provide high-fidelity virtual replicas for network planning, performance evaluation, fault diagnosis, resource optimization, and pre-deployment testing. As Beyond-5G (B5G) and 6G networks become increasingly intelligent, distributed, and heterogeneous, DT-assisted network management has emerged as one of the key research directions for future wireless systems.

Most existing studies along these research directions adopt a \emph{network-centric} perspective, where the communication network itself is the object being twinned. In this paradigm, DTs replicate network topology~\cite{zhou2025cooperative}, radio environments~\cite{wang2025radio}, protocol behaviors~{\cite{traffic_eng_survey}}, traffic dynamics~\cite{lu2025digital}, and resource utilization~{\cite{OptimizingWirelessResource2025}} to better understand and optimize wireless infrastructures. This perspective has substantially advanced network intelligence by enabling proactive resource management, reliability analysis, and virtual experimentation before deployment, thereby reducing operational risks and deployment costs.

Meanwhile, a complementary and increasingly important paradigm has rapidly emerged over the past few years. Rather than treating wireless networks as the \emph{object} of digital twinning, wireless networks are becoming the \emph{subject}. This is, the fundamental infrastructure through which multiple DTs of complex physical systems are constructed, synchronized, and continuously evolved~{\cite{InternetFederatedDigital2024}}. Examples include autonomous transportation~\cite{zhang2026digital}, industrial automation~\cite{ho2025ai}, precision agriculture~\cite{awais2025advancing}, healthcare~\cite{elgammal2025digital}, robotic swarms~\cite{tong2025digital}, and immersive extended reality (XR)~\cite{kuru20256g}, where wireless sensing, communication, edge computing, and artificial intelligence (AI) collectively maintain digital representations of physical environments in real time. In this vision, communication networks no longer simply transport data between applications; instead, they become an active component of the DT ecosystem, enabling continuous interaction between physical and cyber spaces through sensing, learning, simulation, and control.

This evolution fundamentally changes the role of wireless systems in DT-enabled applications. Future communication networks must support not only reliable connectivity, but also semantic information abstraction, distributed intelligence, real-time synchronization, and trustworthy feedback control. Consequently, modern DT systems are increasingly characterized by dynamic operation, where heterogeneous sensing data, AI models, simulation engines, and network control continuously interact in closed loops. Rather than serving as passive virtual replicas, DTs evolve into intelligent cyber-physical agents capable of adapting to environmental changes, coordinating distributed resources, and supporting application-aware decision making across edge-cloud infrastructures.

\begin{figure*}[t]
    \centering
    \includegraphics[scale=0.25]{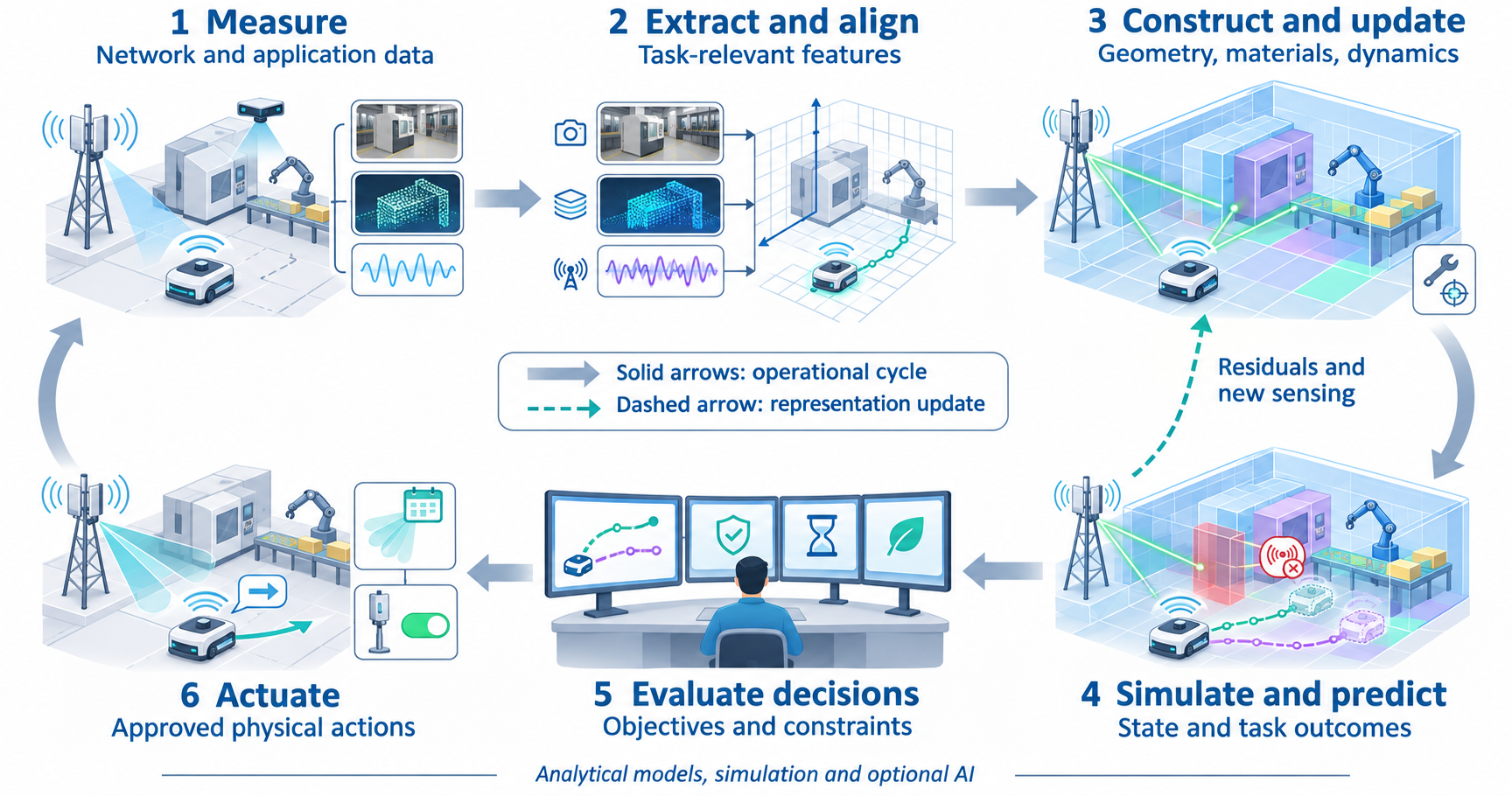}
    \caption{Six-stage recurring DT lifecycle, beginning with measurement and proceeding through feature alignment, twin construction or update, prediction, decision evaluation, and physical actuation. Solid arrows show the operational cycle, and the dashed return path uses residuals and new sensing to revise the digital representation.}
    \label{fig:DTN_design}
\end{figure*}

Although these developments have received growing attention from multiple research communities,
the literature remains highly fragmented (as detailed in Sec.~II). Existing surveys primarily focus on DT-enabled wireless networks, communication technologies for data synchronization, or AI techniques for network optimization individually. Comparatively little attention has been devoted to understanding how wireless infrastructures enable DTs constructed from network data, and how sensing, communication, learning, simulation, and control collectively form an end-to-end DT lifecycle across heterogeneous application domains. As a result, there is currently no unified tutorial that connects these emerging research directions under a common architectural perspective.

This tutorial aims to bridge this gap (as detailed in Fig.~\ref{fig:intro}). Rather than proposing a new DT architecture or a new closed-loop optimization framework, we provide a unified view of the rapidly evolving DT ecosystem from the perspective of next-generation wireless systems. Specifically, we distinguish the traditional \emph{digital twins for wireless networks} paradigm from the emerging \emph{digital twins constructed through wireless networks} paradigm, where network infrastructures serve as the enabling substrate for building, synchronizing, and operating application-aware DT systems. {
Here, application awareness means a physical task determines the modeled state, acceptable uncertainty, update triggers, and decision objective, beyond assigning a service class or a throughput target. For a mobile robot, the DT includes pose, velocity, obstacle geometry, localization covariance, queue backlog, and predicted blockage. Decisions are evaluated by tracking error and collision risk as well as latency. An XR twin additionally tracks head pose, viewport relevance, rendering backlog, and motion-to-photon delay, where pose changes can trigger updates even when average radio throughput is unchanged. In cooperative perception, object tracks, occlusion, and detection uncertainty determine which sensor regions should be transmitted before a fusion deadline. 
Fig.~\ref{fig:intro} illustrates an application-aware DT framework that connects physical systems, shared twin models, and network management to support application-driven optimization. Specifically, PHY-layer modeling (Sec.~V) provides channel and sensing predictions, while resource optimization (Sec.~IV) translates these predictions into decisions on PHY resources, MAC scheduling, association, transport, and application delivery. The management and feedback function (Sec.~VI) supervises these cross-layer operations by monitoring system performance and adjusting configurations to meet application requirements, rather than serving as an additional protocol layer. Measurements and predictions inform management decisions, while candidate configurations and approved actions are fed back to the physical system. The shared twin (Sec.~III) enables this closed-loop interaction, supported by standards, interfaces, and tools (Sec.~VII).
}

Throughout the tutorial, we adopt the six-stage DT lifecycle shown in Fig.~\ref{fig:DTN_design} as an organizational framework that synthesizes recurring functional components across recent literature. 
{
The six stages in Fig.~\ref{fig:DTN_design} are ordered as follows: (1) measure physical and network data; (2) extract, align, and select features; (3) construct or update the twin representation; (4) execute simulation and prediction; (5) evaluate candidate decisions; and (6) apply authorized actions and obtain new measurements. Existing maps and analytical models can provide initial inputs to Stage 3, while measurement processing and feature extraction precede measurement-driven model construction. These stages can be implemented using various approaches, including analytical queueing models, geometric propagation models, channel estimation, and control methods. AI-assisted approaches are particularly beneficial when dealing with heterogeneous observations and unknown residual dynamics.
}

From a system perspective, this lifecycle naturally introduces several emerging research challenges, including semantic-aware data abstraction, cross-domain synchronization, distributed twin orchestration, trustworthy AI, and scalable edge-cloud collaboration. For example, instead of synchronizing raw measurements, future DT systems may exchange semantic representations of traffic conditions, environmental changes, mobility patterns, or application intents, allowing distributed twins to collaboratively optimize system behaviors while significantly reducing communication overhead. Such capabilities are expected to become increasingly important as future AI-native wireless infrastructures evolve toward supporting large-scale, distributed, and application-centric DT ecosystems.

Accordingly, this tutorial is organized around the following three themes that characterize the emerging evolution of DT-enabled wireless networks, as we illustrate in Fig.~\ref{fig:intro}.

\begin{itemize}

\item \textbf{DTs Constructed through Wireless Networks.}
We first distinguish the conventional paradigm, where communication networks themselves are digitally twinned, from the emerging application-centric paradigm, where wireless infrastructures provide the sensing, communication, computing, and synchronization capabilities required to construct and maintain DTs of external physical systems. This perspective broadens the role of future wireless networks from communication platforms to fundamental enablers of cyber-physical intelligence.

\item \textbf{Dynamic and AI-Native DT Operation.}
We review recent advances that enable DTs to evolve continuously through online synchronization, distributed AI, semantic information processing, edge-cloud collaboration, and adaptive orchestration. Rather than static virtual replicas, modern DTs operate as continuously evolving systems capable of learning from and responding to changing physical environments.

\item \textbf{Closed-Loop DT Lifecycle.}
Instead of viewing modeling, communication, learning, simulation, and control as isolated research topics, we organize this tutorial using a unified closed-loop DT lifecycle. 
{
We distinguish two coupled cycles. The operational control cycle measures the system, estimates its current state, predicts consequences, applies an action, and measures the response. The representation-update cycle uses prediction residuals and new sensing to revise the model itself, including geometry, object identities, materials, propagation parameters, or learned weights, and validates the revision before use. Physical-to-DT and DT-to-physical are directions of information flow within these cycles, rather than two independently complete control loops.
}

\end{itemize}

The remainder of this tutorial is organized as follows, and the link between sections \ref{sec3}-\ref{sec7} is shown in Fig.~\ref{fig:intro}. Section II reviews the existing literature on DTs for wireless networks and positions this tutorial relative to prior surveys. Section III presents the fundamental principles of digital twin network (DTN) creation, including twin modeling, inter- and intra-twin operations, and lifecycle management. Section IV discusses resource optimization and semantic-driven DTNs, highlighting application-aware design and intelligent resource orchestration. Section V focuses on physical-layer optimization techniques as the pillar for DTNs, including sensing, communication, and wireless environment modeling. Section VI examines DTN-guided network management and feedback optimization through closed-loop control and adaptive decision-making. Section VII reviews ongoing standardization efforts, open-source platforms, and enabling tools for DTN deployment. Finally, Section VIII concludes the tutorial and outlines key research challenges and future directions for application-aware, AI-native, and closed-loop DTNs.

\section{Related Work and Positioning}

DT has rapidly evolved into a central architectural concept for beyond-5G and 6G wireless systems. A growing body of survey literature has examined the integration of DT technologies into communication networks from architectural, AI-driven, and edge-centric perspectives. The seminal survey by Khan \textit{et al.} \cite{Khan2022DTWireless} established a comprehensive taxonomy for wireless digital twins, detailing system architecture, synchronization mechanisms, modeling approaches, and open research challenges. This work laid the conceptual foundation for DT-enabled wireless networks. Tang \textit{et al.} {\cite{tang2022survey}} extended this direction by proposing Digital Twin Edge Networks (DITEN), emphasizing the convergence of DT with mobile edge computing and network slicing toward 6G. From the industrial perspective, Lin \textit{et al.} {\cite{lin20236g2}} introduced the vision of 6G Digital Twin Networks (DTNs), discussing architectural design and representative use cases and tools across vertical domains. Subsequent surveys have increasingly focused on AI-native DT systems. For instance, Tao \textit{et al.} {\cite{tao2024wireless}} examined generative AI as an enabling technology for wireless network digital twins, highlighting data-driven modeling and adaptive learning capabilities. Sheraz \textit{et al.} \cite{Sheraz2024AIDTN} surveyed AI-enabled DTs in 6G, emphasizing intelligent orchestration and automation. {
Sai \textit{et al.} 
}\cite{Sai2024DTNLandscape} provided a broader architectural overview of DT networks with attention to privacy, security, and communication models. More recently, a few works have expanded the discussion toward large-scale multi-domain ecosystems. The survey in \cite{DTN2025E2E} presents an end-to-end view of network DTs across heterogeneous domains, including radio access network (RAN), core, edge-cloud, and non-terrestrial networks. Meanwhile, \cite{RayTracing2025} explores multi-task adaptive ray tracing as a key enabler for 6G DT networks, focusing on channel modeling fidelity and environment-aware simulation.
\cite{Sai2024DTNLandscape} provided a broader architectural overview of DT networks, highlighting key design considerations related to privacy, security, communication models, and interoperability. More recently, several studies have extended this discussion toward large-scale, multi-domain DT ecosystems. The survey in \cite{DTN2025E2E} presents an end-to-end perspective on network DTs spanning heterogeneous domains, including radio access networks (RANs), core networks, edge-cloud infrastructures, and non-terrestrial networks. Meanwhile, \cite{RayTracing2025} investigates multi-task adaptive ray tracing as a key enabler for 6G DT-enabled networks, focusing on improving channel modeling fidelity and environment-aware simulation while balancing computational complexity and modeling accuracy. %

{
Prior surveys and vision papers already discuss feedback and application support. In particular, \cite{Alkhateeb2023RealTime} describe proactive handover, edge-service migration, and video preloading using environment-aware predictions. Our tutorial complements this work by connecting a state-space construction example to resource allocation, physical reconstruction, and management. We use application awareness as a design criterion rather than claiming that application-driven twinning or cross-layer control is new. Table~\ref{tab:comparison_surveys} summarizes the topics discussed, not a ranking of the completeness of prior work.
}

\begin{table*}[t]
\centering
\caption{Comparison of Existing DT-for-Wireless Tutorials.}
\label{tab:comparison_surveys}
\resizebox{\textwidth}{!}{\begin{tabular}{lcccccccc}
\hline
\textbf{Tutorial/Survey} & \textbf{Arch.} & \textbf{AI} & \textbf{Edge} & \textbf{Closed-Loop} & \textbf{App.-Aware} & \textbf{Cross-Layer} & \textbf{Semantic / ISAC} & \textbf{Standards \& Tools} \\
\hline
Khan et al. (2022) \cite{Khan2022DTWireless} 
& \checkmark & Partial & Partial & Conceptual & \textemdash & \textemdash & \textemdash & Partial \\

Tang et al. (2022) {\cite{tang2022survey}} 
& \checkmark & Partial & \checkmark & \textemdash & Partial & \textemdash & \textemdash & Partial \\

{Lin et al. (2023)} {\cite{lin20236g2}} 
& \checkmark & Partial & \checkmark & Conceptual & Partial & \textemdash & \textemdash & \checkmark \\

{Alkhateeb et al. (2023) \cite{Alkhateeb2023RealTime}} & {Partial} & {Partial} & {Partial} & {\checkmark} & {\checkmark} & {\checkmark} & {Partial} & {Partial} \\

{Tao et al. (2024)} {\cite{tao2024wireless}} 
& \checkmark & \checkmark & Partial & \textemdash & \textemdash & Partial & \textemdash & \textemdash \\

Sheraz et al. (2024) \cite{Sheraz2024AIDTN} 
& \checkmark & \checkmark & \checkmark & Partial & Partial & Partial & \textemdash & Partial \\

Sai et al. (2024) \cite{Sai2024DTNLandscape} 
& \checkmark & Partial & Partial & \textemdash & \textemdash & \textemdash & \textemdash & \textemdash \\

Tran et al. (2025) \cite{DTN2025E2E} 
& \checkmark & Partial & \checkmark & Conceptual & Partial & Partial & \textemdash & Partial \\

Yu et al. (2025) \cite{RayTracing2025} 
& \checkmark & Partial & Partial & \textemdash & \textemdash & Partial & \textemdash & \textemdash \\

\textbf{This Tutorial} 
& \checkmark & \checkmark & \checkmark & \checkmark & \checkmark & \checkmark & \checkmark & \checkmark \\

\hline
\end{tabular}}
\end{table*}

\section{Fundamental Design of DTN creation, Inter/Intra Twin Operations}
\label{sec3}

\subsection{Motivation and Basic Problems}

A DTN is a collection of interconnected, task-oriented DTs deployed across a physical network, where each twin models a specific subsystem, region, protocol layer, or network function. These twins jointly represent the network’s evolving dynamics by monitoring heterogeneous states, processing multi-modal measurements, and supporting closed-loop control and optimization. Rather than existing in isolation, individual twins in a DTN collaborate through inter- and intra-twin operations, enabling knowledge sharing, model reuse, and coordinated adaptation across spatially distributed and functionally diverse network components.

A fundamental and long-standing problem in the field lies in how to create the DTN from scratch in a principled, scalable, and theoretically grounded manner. Unlike application-specific or task-driven models as in the field of manufacturing or urban planning, DTN creation aims to establish a persistent and reusable virtual counterpart of an ever-changing network system that can faithfully track channel dynamics and support control decisions across heterogeneous deployments. This problem is particularly significant because the quality of all subsequent twin operations, such as model reuse, transfer, merging, critically depends on the fidelity, structure, and adaptability of the initially constructed twin. However, existing works often treat DT construction as an ad-hoc modeling step tied to specific tasks or datasets, lacking a unifying mathematical backbone that clearly defines what it means to map a physical network into a DT.

From a theoretical perspective, DTN creation is fundamentally a problem of modeling, estimation, and control, demanding a rigorous mathematical foundation rather than heuristic design choices. Building upon the foundational concepts of NASA digital twins~\cite{allen2021digital,kapteyn2021probabilistic}, the DTN creation problem can be cast into a state-space model (SSM). This formulation provides a compact yet expressive representation for capturing three co-evolving processes in a unified framework: (i) the intrinsic dynamics of the physical network, (ii) the internal data processing and inference flow of the twin model, and (iii) the higher-level distributed mapping and optimization procedures that continuously refine the twin. By adopting an SSM abstraction, DTN creation can be studied using well-established system-theoretic notions such as stability, observability, identifiability, and controllability, elevating DTNs from collections of loosely coupled learning models to formally analyzable dynamic systems.

Consider a generic continuous-time nonlinear SSM that describes the physical wireless network dynamics,
\begin{equation}
    \dot{x}(t) = f(x(t), u(t)) + w(t),
    \label{eq:generic_state_update}
\end{equation}
\begin{equation}
    y(t) = h(x(t), u(t)) + v(t),
    \label{eq:generic_output_equation}
\end{equation}

\noindent where $x(t)$ denotes the latent system state, $u(t)$ represents control inputs, $y(t)$ is the observable output, and $w(t)$ and $v(t)$ capture process and measurement noise, respectively. In the DTN context, the state $x(t)$ corresponds to the high-dimensional and spatially distributed network conditions, such as traffic demand, mobility patterns, topology evolution, and channel states. The outputs $y(t)$ aggregate heterogeneous multi-modal measurements collected from access points, sensors, and protocol stacks, while $u(t)$ reflects network control actions including scheduling, routing, and resource allocation.

{
Consider a mobile robot served by a base station (BS), which maintains a queue of downlink data destined for the robot. Under a discrete-time model with sampling interval $\Delta$, the system state at time step $n$ is defined as $x_n=(q_n,p_n,\nu_n,g_n)$, where $q_n$ denotes the BS queue backlog (in bits), $p_n$ and $\nu_n$ represent the robot's pose and velocity, respectively, and $g_n$ denotes the downlink channel power gain. The control action is $u_n=(b_n,P_n,c_n)$, where $b_n$ is the allocated bandwidth, $P_n$ is the BS transmit power, and $c_n$ is the robot's motion command. A concrete transition is
\begin{equation}
\begin{split}
&q_{n+1}=[q_n-\Delta R(g_n,b_n,P_n)]^++a_n,\\
&p_{n+1}=p_n+\Delta \nu_n,\quad
v_{n+1}=v_n+\Delta c_n+\eta_n,\\
&R(g,b,P)=b\log_2(1+Pg/(N_0b)).
\end{split}
\label{eq:queue_example}
\end{equation}
Here $a_n$ is the number of arriving bits per interval, $N_0$ is noise power spectral density, and $\eta_n$ is motion disturbance. The gain map $G$ depends on pose and scene state $m_n$, with calibrated propagation parameters $\theta_g$ and residual $\xi_n$. Scene geometry and material changes update $m_n$ or the representation of $G$. This illustrative rate omits inter-cell interference; a deployed model replaces it by measured goodput or an interference-aware expression. Queue counters, odometry, and pilot-based channel measurements give $y_n=(q_n,p_n,g_n)+v_n^{\rm obs}$, an explicit observation map $h$; unmeasured velocity is estimated. The right-hand sides define a sampled counterpart of $f$ in Eq.~\eqref{eq:generic_state_update}.  Arrival statistics, motion response, and gain evolution are calibrated from data; queue conservation and known bandwidth limits can remain fixed.

At each site, local trajectory fitting estimates these uncertain parameters; aggregation shares compatible parameters, and fusion combines aligned predictors. A fitted model is then held fixed while a scheduler evaluates bandwidth and motion choices. If predicted blockage reduces $g_n$, the queue may delay a robot observation beyond its control deadline. The controller can reserve bandwidth, change association or edge placement, or slow the robot, choosing the action with acceptable tracking risk. 
In essence, the DT provides an approximation $(\hat f,\hat h)$ of the physical dynamics $(f,h)$ using analytical structure, calibrated parameters, learned components, or a combination. 
}

\subsection{State of the Art and Gaps}

Based on this problem statement, this section reviews prior studies related to DTN creation and lifecycle management, with a focus on three aspects: direct DT mapping and creation, inter- and intra-twin operations for reuse, and post-adjustment, validation, and backward optimization.

\subsubsection{Direct Mapping and Creation of DTN}

DTN have been widely recognized as a fundamental enabler for next-generation networking systems. A large body of work highlights their role in enhancing the planning, deployment, and optimization of 5G and beyond networks, particularly in improving coverage, spectral efficiency, and end-to-end throughput~{\cite{nguyen2021digital,sandbox,zhang2025synergizing2,tang2025research,tao2024ran,li2024intelligent,tan2025energy}}. 
The architectural requirements and system-level considerations for effective DTN deployment are discussed in~{\cite{mashaly2021connecting}}, emphasizing the importance of continuous data acquisition, real-time analytics, and closed-loop control. 
From a modeling perspective, graph-based learning techniques have been introduced to improve DT expressiveness, such as employing graph neural networks (GNNs) to model network topology and predict per-slice traffic demand for efficient network slicing~\cite{graphneuralnetwork}.

In application-driven domains, DTNs have been extensively studied in vehicular and V2X networks. Works such as~\cite{intelligentdigitaltwin,Ding2024digital} demonstrate how DTs can support proactive resource management, cooperative perception, and reliability enhancement under high mobility. 
Despite these advances, most existing studies primarily focus on \emph{how to use} DTs for network optimization, rather than \emph{how to construct} DTs that faithfully represent the physical infrastructure and its evolving dynamics. The twin creation process, encompassing data collection, physical-to-digital mapping, and continuous refinement, is often abstracted away or treated as given~{\cite{10375758,Khan2022DTWireless,Yang2024ICASSP}}.
Only recently have a limited number of works started to address DT construction from scratch. These efforts explore systematic pipelines for sensing, differentiable modeling, and online refinement of digital twins using real-world measurements~{\cite{villa2024colosseum,10255711,10619056}}. While these approaches provide important foundations for direct DT creation, they typically rely on centralized processing or task-specific designs and do not fully address scalability, heterogeneity, or reuse across multiple DT instances.

\subsubsection{Inter- and Intra-Twin Operations and Model Reuse}

Beyond initial creation, practical DTNs require mechanisms for inter- and intra-twin operations, including reuse, transfer, merging, and splitting of task-oriented twins. Distributed learning has emerged as a critical enabler in this context, allowing large volumes of network data collected across access points and sensors to be processed in parallel while reducing communication overhead. The importance of distributed intelligence for DT systems has been formalized in ITU standards~{\cite{ITUT2022Y3090,ITUT2023Y3091}}.

Federated learning (FL), in particular, has been widely studied as a communication-efficient distributed optimization framework~\cite{niknam2020federated,zhao2021federated,9210812,11270936}. These works address key challenges such as statistical heterogeneity, partial client participation, and convergence guarantees. Building on these foundations, prior efforts have explored federated DT construction through vertical and horizontal mapping strategies, enabling DTs to be built and refined collaboratively across distributed sites~{\cite{10619056}}. However, most existing FL-based approaches focus on training a single global model and provide limited support for structured inter-twin reuse or task-level composition.

In parallel, multi-modal learning has become a powerful paradigm in AI/machine learning (ML), exploiting complementary information across heterogeneous data sources to improve model robustness and generalization~\cite{huang2021makes}. Transformer-based architectures have been shown to be particularly effective for data-level multi-modal fusion~\cite{xu2023multimodal}, while other works demonstrate gains from integrating cross-domain signals in applications such as remote sensing and perception~\cite{hong2020more}. At the task level, self-supervised and multi-task multi-modal learning has been explored to generate modality-specific and transferable representations~\cite{yu2021learning}.  
{
These methods offer complementary options: concatenation with a calibrated predictor is a simple baseline, Bayesian filtering exposes uncertainty, FL shares compatible parameters without raw data transfer, and mixture models allow context-specific predictors. 
}

\subsubsection{Post-Adjustment and Validation}

After deployment, DTNs must support post-adjustment operations such as validation, rollback, and selective removal of outdated or undesired contributions. Traditional DT studies largely assume monotonic model growth and focus on forward refinement through continual learning. In contrast, backward operations remain underexplored despite their importance for long-lived, regulated, and privacy-sensitive network environments.

Machine unlearning has emerged as a relevant line of research, aiming to remove the influence of specific data points or clients from trained models~\cite{cao2015towards,bourtoule2020machineunlearning}. This capability is essential for privacy protection, compliance with data retention policies, and mitigation of data poisoning attacks. Federated learning further complicates unlearning due to the distributed nature of data contributions, leading to the development of federated unlearning techniques~\cite{liu2020federaser,Fraboni2023sifu,halimi2023federatedunlearningefficientlyerase}. However, most unlearning methods are designed for classification-centric ML models and do not account for tightly coupled spatio-temporal dynamics or continuous-valued predictions typical in DTNs.

{
Recent studies connect DTs with privacy-aware learning and federated unlearning~{\cite{yuan2024iot,daluwatta2024dt}}. The untwinning case study in~\cite{zhang2026network} extends this direction to removal of scenario-dependent twin contributions. Full DT reconstruction remains a reference baseline, where checkpoint replay and existing federated unlearning methods offer alternatives with different storage, computation, and auditing requirements.
}

\subsection{Challenges and Solutions}

\subsubsection{Key Challenges}

\paragraph{{C1. Nonconvex and partially observed identification}}
The core of DTN creation is the identification of a global continuous-time nonlinear SSM from distributed, noisy, and heterogeneous observations. Formally, learning the optimal dynamics $(\hat f,\hat h)$ in
\begin{equation}
    \dot{\hat{x}}(t)=\hat f(\hat{x}(t),u(t)), \qquad
    \hat{y}(t)=\hat h(\hat{x}(t),u(t))
\end{equation}
{
can involve nonconvex system identification when neural or nonlinear parameterizations are employed. The difficulty depends on the model class, observation process, and loss function. While linear-in-parameter identification can be tractable under suitable excitation and observability conditions, partial observability, asynchronous sampling, and coupled parameters introduce substantial challenges in wireless network settings.
}

\paragraph{C2. Multi-modal Integration and Twin Reuse}
DTNs ingest heterogeneous sensing modalities with incompatible state spaces and temporal resolutions~{\cite{11309199}}. Let $y^{(k)}(t)=\{y^{(k)}_1(t),\dots,y^{(k)}_m(t)\}$ denote the $m$ modalities observed at site $k$. Constructing a reusable twin requires learning a shared latent embedding $\phi(t)\in\mathbb{R}^d$ such that
\begin{equation}
    \phi(t)=\mathcal{F}\big(y^{(k)}_1(t),\dots,y^{(k)}_m(t)\big),\label{eq:feature_fusion}
\end{equation}
preserves semantically relevant dynamics across tasks. Naive fusion operators, such as concatenation or averaging, often lead to task interference and negative transfer, limiting the effectiveness of inter-twin reuse and model composition.

\paragraph{C3. Backward Optimization and Validation after Creation}
DTN construction must support backward operations such as validation, rollback, and selective removal of obsolete components. 
Let $\Theta^\star$ denote the converged DTN model. When part of the training data or a local DT must be removed, a naive solution would require recomputing
\begin{equation}
    \Theta^\star = \arg\min_{\Theta} \sum_{k=1}^K \omega_k F_k(\Theta),
\end{equation}
from scratch, which is infeasible in real-time settings. 
{
Specifically, $\Theta=(\theta_f,\theta_h,\theta_e,\theta_{\mathcal F})$ is the trainable dynamics, observation, encoder, and fusion parameters present in the chosen implementation. 
For the queue example, known queue conservation is fixed while an arrival predictor or channel residual can be learned. Define $\mathcal T_k=[t_{k,0},t_{k,1}]$ as site $k$'s observation interval, $\ell_k$ as a nonnegative prediction loss (for example, weighted squared error with each output scaled by a stated physical reference), and $F_k$ as its integral over this interval as specified below. Nonnegative aggregation weights $\omega_k$ sum to one and may represent sample fractions; their choice must be stated. $F=\sum_k\omega_kF_k$ fits model parameters and is distinct from the feature fusion map $\mathcal F$ in Eq.~\eqref{eq:feature_fusion}.
}
Furthermore, due to coupling constraints among DTs, local rollback operations can propagate errors globally, challenging both model consistency and stability.

\subsubsection{Solution I: Multi-modal FL for SSM Construction}

{
To distribute data processing and model fitting, DTN creation can be implemented as federated optimization over compatible SSM parameterizations. This changes where optimization runs; it does not eliminate nonconvexity or guarantee a global optimum.
} Each DT instance $k$ performs local trajectory-aware optimization of
\begin{equation}
    F_k(\Theta)=\int_{\mathcal{T}_k} 
    \ell_k\big(\hat{y}_{\Theta}(t;k), y^{(k)}(t)\big)\,\mathrm{d}t,
\end{equation}
using locally available multi-modal observations and control inputs. These local updates are periodically aggregated by a global coordinator:
\begin{equation}
    \min_{\Theta}\; F(\Theta)=\sum_{k=1}^K\omega_k F_k(\Theta),
\end{equation}
via communication-efficient parameter fusion.  

{
Note that local encoders can reduce the data sent to a coordinator, although communication per round still scales with the number of participating sites and transmitted parameters. Convergence claims require an explicit algorithm and assumptions. For example, FedProx analyzes heterogeneous federated optimization under regularity and bounded-dissimilarity conditions~\cite{Li2020FedProx}, but it does not establish convergence for arbitrary changing physical systems. Unaligned dynamics may require personalized models, clustered aggregation, or local analytical estimators. Training loss convergence also does not imply closed-loop stability.
}

\subsubsection{Solution II: Inter- and Intra-Twin Operations via Structured State-Space Fusion}

Beyond initial construction, DTNs must support dynamic reuse and evolution of task-oriented twins. Structured state-space fusion provides a principled mechanism for such inter- and intra-twin operations. Specifically, rather than enforcing a single homogeneous model, the global dynamics are expressed as a context-dependent mixture of local operators:
\begin{equation}
    \dot{\hat{x}}(t)=
    \sum_{k=1}^K \pi_k\!\big(z(t)\big)\,
    f^{(k)}_{\theta_k}\!\big(\hat{x}(t),u(t)\big),
\end{equation}
where $z(t)$ encodes environmental or operational context. In this formulation, intra-twin operations correspond to refining local parameters $\theta_k$ using site-specific data, while inter-twin operations correspond to adapting the aggregation weights $\pi_k$ or transferring model components across twins. {
Before evaluating this mixture, local states must be mapped to a common coordinate system, units, state dimension, and time base. With $x=A_k(x_k)$ and a differentiable invertible alignment on the shared subspace, the compatible operator is $\bar f_k(x,u)=D A_k(A_k^{-1}x)f_k(A_k^{-1}x,u)$. The operators written as $f^{(k)}_{\theta_k}$ in this equation denote these aligned operators.
The gating weights satisfy $\pi_k\geq0$ and $\sum_k\pi_k=1$.
}

This structured fusion enables controlled merging, splitting, and transfer of DTs while preserving continuous-time semantics. Importantly, it supports the synthesis of new task-oriented twins from existing ones without requiring direct physical-to-twin instantiation, significantly reducing cold-start costs and accelerating DTN expansion across large deployments.

\subsubsection{Solution III: Digital Untwinning for Backward Optimization}

Digital untwinning~\cite{zhang2026network} introduces a systematic mechanism for backward optimization in DTNs by selectively removing the influence of specific DTs or data contributions. Let $\mathcal{S}\subseteq\{1,\dots,K\}$ denote the set of DTs to be removed. Instead of full retraining, untwinning solves a constrained correction problem:
\begin{equation}
    \min_{\Theta}\;
    \sum_{k\notin\mathcal{S}}\omega_k F_k(\Theta)
    \quad
    \text{s.t. }\|\Theta-\Theta^\star\|\le \epsilon,
\end{equation}
{
where $\Theta^\star$ is the pre-removal model and $\epsilon$ is a trust-region radius. This constraint limits displacement from the old twin model; it does not certify agreement with retraining on retained data. That separate property requires comparison with a retained-data retraining reference $\Theta^\star_{-\mathcal S}$, for example by evaluating prediction disagreement on retained and removed domains. A tight radius may prevent adequate removal, and the radius must be relaxed or the model retrained if the audit fails.
}

Practically, untwinning mechanisms rely on sensitivity-aware rollback points, dependency analysis among interconnected DTs, and fast remapping operators that restore model consistency. By treating backward operations as first-class primitives rather than exceptions, digital untwinning enables long-lived DTNs to evolve under changing data availability, regulatory constraints, and operational objectives {
with stability and removal quality requiring separate validation.
}

\subsubsection{Solution IV: Hybrid Twinning Stack with Inner–Outer Loop Adaptation}

A long-lived DTNs requires jointly handling fast-varying, interpretable system parameters \textit{and} slowly evolving, highly nonlinear behaviors under real-time constraints. Pure data-driven models struggle with stability, sample efficiency, and adaptation speed, while purely model-based approaches lack expressive power when system dynamics deviate from linear assumptions. To address this gap, the promising solution is to develop a hybrid DT stack that decomposes learning and adaptation across two tightly coupled loops. This structure can be captured as:
\begin{equation}
    y_t = \boldsymbol{\theta}_t^\top \mathbf{x}_t + f_{\text{ML}}(\mathbf{x}_t) + \varepsilon_t,
\end{equation}
where $\boldsymbol{\theta}_t$ denotes linear coefficients such as channel gains, queue slopes, or damping parameters that can be efficiently updated in real time. In contrast, $f_{\text{ML}}(\cdot)$ represents nonlinear dynamics, higher-order interactions, and long-range dependencies learned using ML models at a slower timescale. $\varepsilon_t$ captures measurement noise, such as sensor noise or packet-level randomness. This decomposition enables fast, stable inner-loop adaptation while preserving expressive power and generalization through outer-loop learning.

Specifically, the \textbf{inner loop} focuses on real-time estimation of those linear or linear-in-parameter components using lightweight adaptive estimators including  recursive least square (RLS) estimation or Kalman filtering. {
Their guarantees depend on assumptions: RLS needs informative excitation and appropriate forgetting and noise conditions, while Kalman filtering requires a specified noise model and suitable observability/detectability conditions. These estimators update $\theta_f$, $\theta_h$, or latent states in the SSM, and a learned residual augments $\hat f$ or $\hat h$. Controller stability additionally requires checking delay, model mismatch, and the chosen feedback law, but it is not implied by estimator convergence.
} The \textbf{outer loop} operates at a slower timescale and leverages the aforementioned learning-based models to capture nonlinear residuals, cross-layer couplings, and long-term multi-twin structural patterns that are difficult to encode analytically. By periodically refining feature representations, twin model priors, or parameter constraints supplied to the inner loop, the outer loop enhances expressiveness without violating real-time requirements.

This separation of concerns transforms DTN creation into a structured, multi-timescale parameterized optimization process, where fast adaptive estimation ensures local fidelity while slower data-driven updates improve global accuracy and generalization. {
This hybrid decomposition is a design option whose responsiveness, estimation error, and closed-loop behavior must be assessed for the deployment. 
}

\subsection{Use Cases and Examples}

\begin{figure}[t]
    \centering
    \includegraphics[scale=0.18]{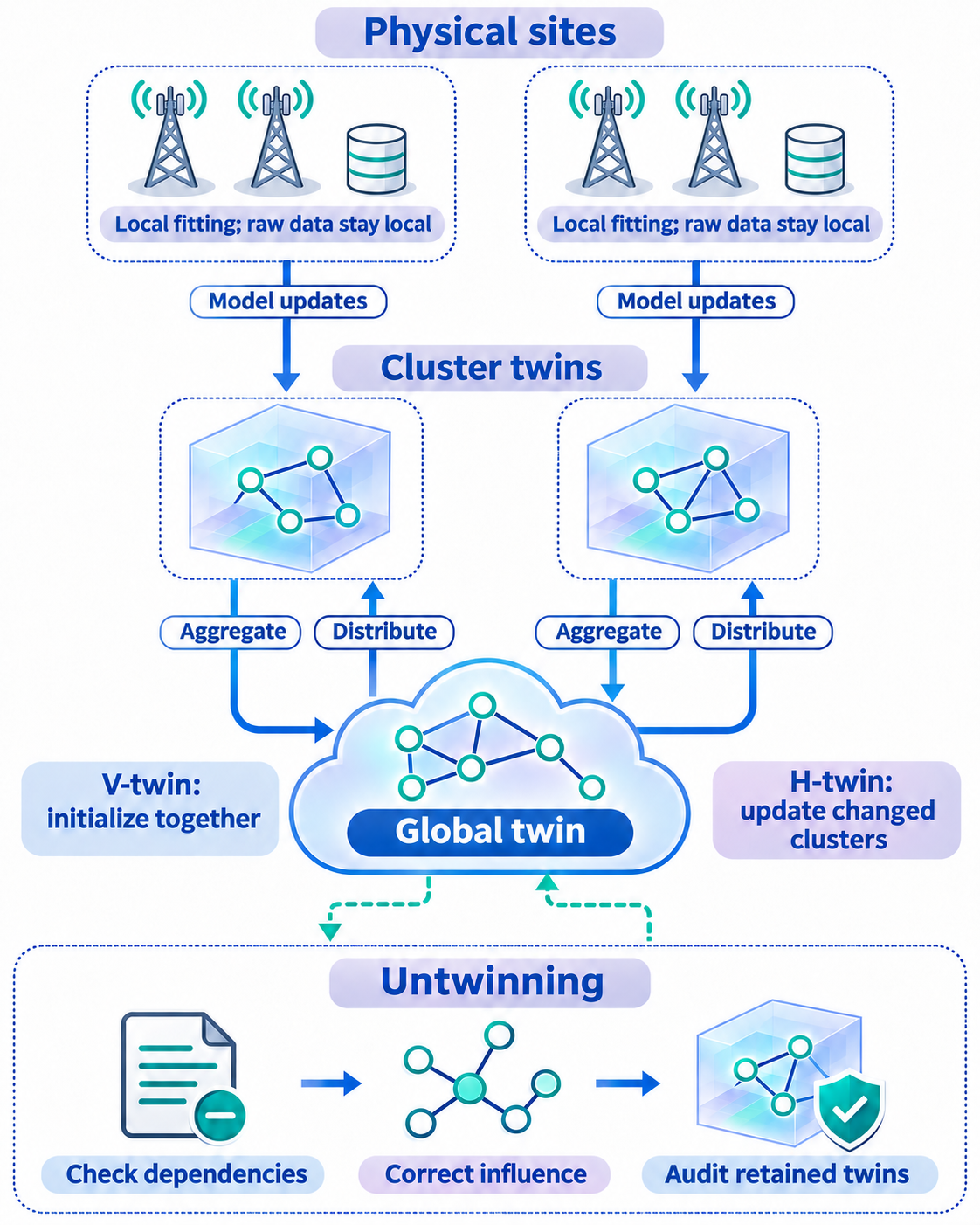}
    \caption{Hierarchical twin construction and untwinning framework. Sites retain raw data and share model updates; cluster twins aggregate into a global twin. V-twinning initializes jointly, while H-twinning updates changed clusters. Untwinning corrects mapped influence and audits retained twins without removing physical sites.}
    \label{fig:twin_sites}
\end{figure}

\begin{figure}[htbp!]
    \centering
    \includegraphics[scale=0.3]{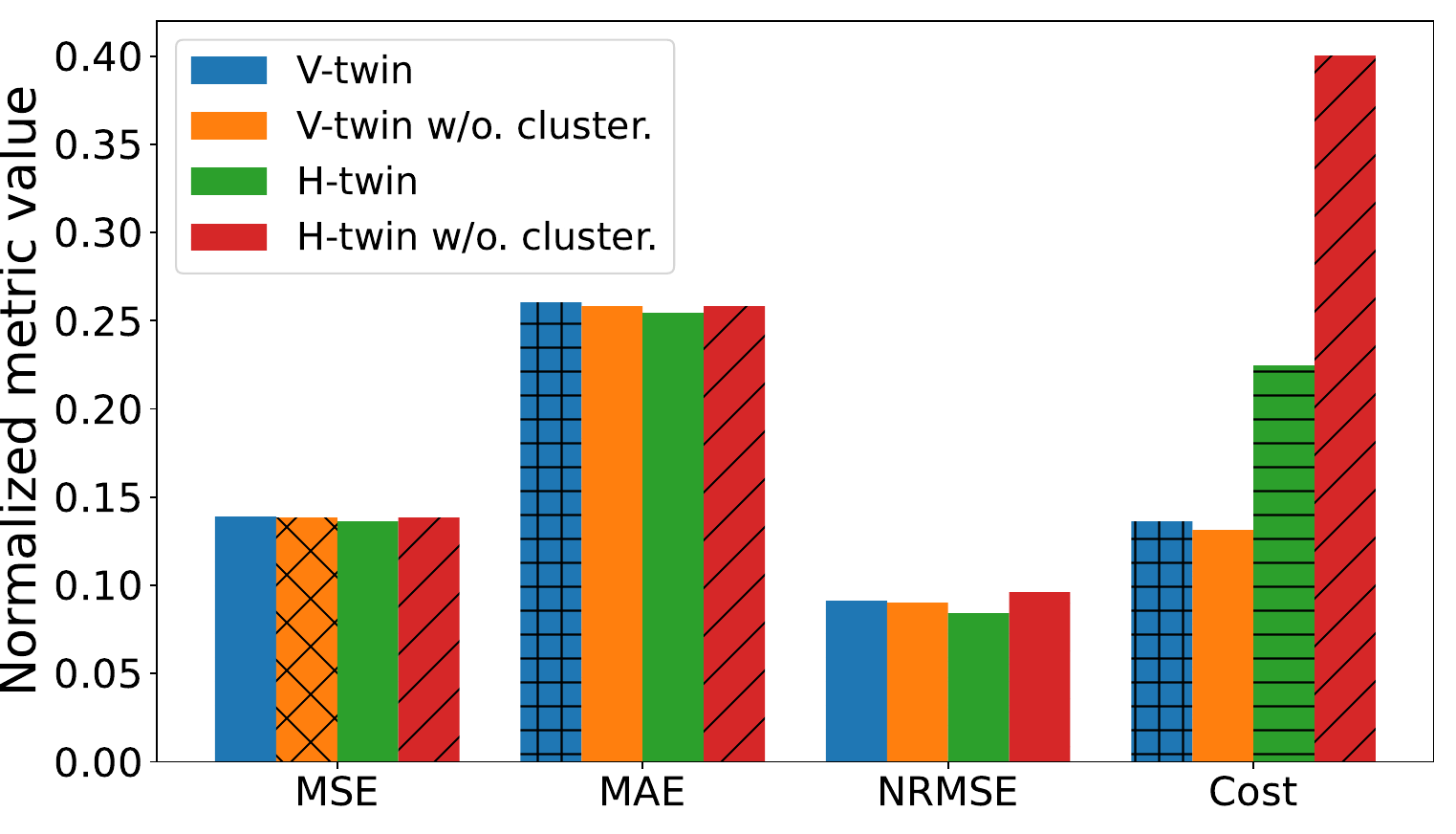}
    \caption{\small DTN construction performance over time-series data.}
    \label{fig:vh}
\end{figure}

{
The following case studies (depicted in Fig.~\ref{fig:twin_sites}) illustrate an integrated solution for the forward construction and backward removal of DTNs.
}
Specifically, {\cite{10619056,zhang2024digital2}} developed federated construction methods of DTN tailored specifically for wireless networks and applied it to time-series data traffic forecasting. As proposed in~{\cite{10619056}}, it 
validated that the DTN can be modeled through joint vertical mapping (\textit{V-twin}) and horizontal mapping (\textit{H-twin}) process. Specifically, V-twin initializes a global twin model from the historical network traffic of virtualized clusters, while H-twin is implemented with an asynchronous fashion that dynamically updates distributed components in response to network changes. The resulting self-evolved twins {
can be applied
} for collaboratively forecasting precise time-series data, such as heterogeneous wireless traffic in cellular networks. 
{
In Fig.~\ref{fig:vh}, MSE, MAE, and NRMSE summarize forecasting error, while Cost summarizes construction or update overhead rather than prediction loss. It also distinguishes initial mapping time for V-twinning from update communication overhead for H-twinning~{\cite{10619056}}. The 43.8\% reduction applies to H-twin relative to its non-clustered baseline. Since the supplied normalized cost do not specify a common physical denominator, they should only be compared within each phase, not interpreted as seconds or a universal compression ratio.
}

\begin{figure}[htbp!]
    \centering
    \includegraphics[scale=0.4]{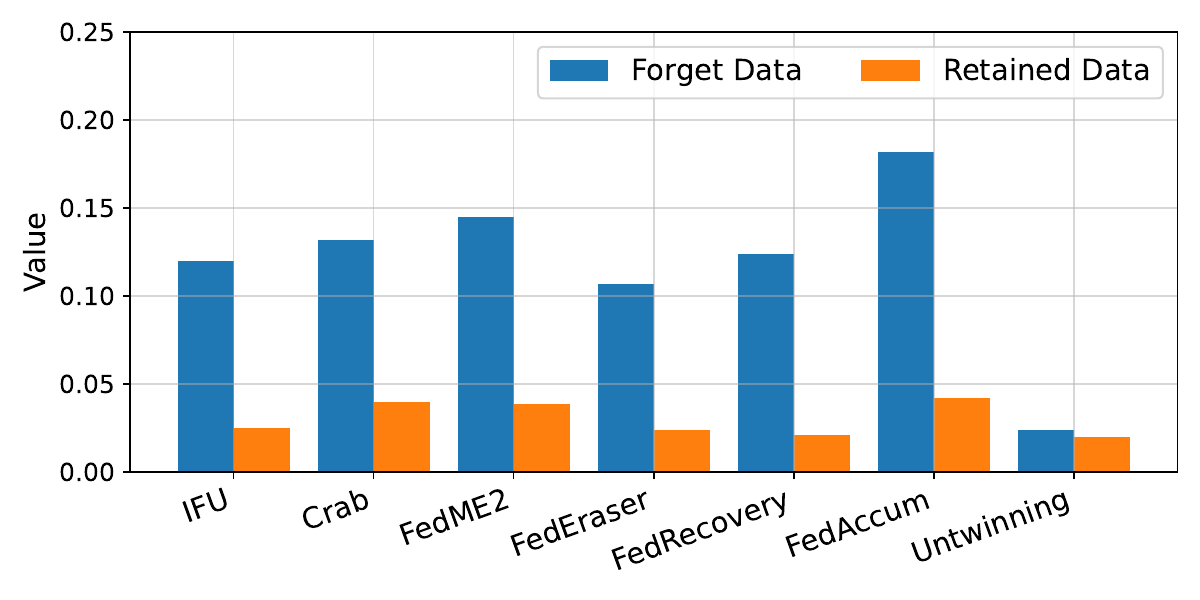}
    \caption{{\small Vehicle-speed prediction-error difference from map-from-scratch after an untwinning request, on forget and retained data. Lower is closer to the retraining reference.
    }}
    \label{fig:untwin}
\end{figure}

For the backward process, ~\cite{zhang2026network} introduced a prototype digital untwinning framework %
using a real-world traffic data scenario from a segment of the I-15 freeway in Utah, U.S. 
The collected dataset\footnote{The dataset has been uploaded to the Department of Transportation and is available upon request~\cite{udot2025traffic}.} aggregates traffic data from inductive loop detectors installed along the freeway.
The framework consists 
of two main modules: single-request untwinning (SRU) and parallel-request untwinning (PRU). Both modules analyze dependencies among local domains orchestrated by a global DT in a federated architecture.
Single-request untwinning performs backward optimization for individual untwinning operations when detecting malicious entities or handling privacy-preserving requests. PRU, on the other hand, addresses large-scale networks in which many interconnected local DTs receive simultaneous untwinning requests.

Fig.~\ref{fig:untwin} evaluates the performance of untwinned models produced by our methods, compared with the ground-truth map-from-scratch twin. The results demonstrate that the untwinning method consistently achieves the lowest prediction difference in vehicle speed predictions.
{
Note that an untwinning request removes a site's or scenario's learned influence, not merely its input stream or the physical device. Correlated neighbors can retain that influence, so dependency checks select affected models for correction and retained twins are revalidated. In Fig.~\ref{fig:untwin}, Value denotes prediction-error difference relative to the map-from-scratch twin reference for vehicle-speed forecasting, reported separately on forget and retained data. Smaller values mean closer agreement with retraining, not a higher application utility or proof of privacy.
}
However, these prior efforts do not account for the unique properties of federated continuous learning cycles or the security implications of hierarchical DTN ecosystems. Therefore, {future works
can build upon these results to develop a unified backward optimization framework that integrates DT-level and telemetry-level backward learning in two stages, along with a concrete post-auditing mechanism under adversarial conditions.}

\subsection{Summary of Learned Lessons}
Several key lessons emerge regarding the creation and operation of DTNs. First, effective DTNs cannot be built as purely data-driven or task-specific artifacts. A principled mathematical foundation, such as state-space modeling, is essential to ensure fidelity, interpretability, and long-term evolvability of twins that represent complex, dynamic network systems. Direct mapping from physical networks to twins must therefore be treated as a core design problem rather than an implicit preprocessing step.

Second, scalability and heterogeneity fundamentally shape DTN design. {
Distributed and federated learning are useful options under bandwidth or data-locality constraints, while sufficient backhaul and a tractable analytical model may instead favor centralized estimation.
} Similarly, multi-modal learning must go beyond naive fusion and explicitly account for modality imbalance, semantic alignment, and reuse across tasks to avoid negative transfer in inter- and intra-twin operations.

Finally, DTNs are inherently long-lived systems that must support backward operations such as validation, rollback, and selective removal of outdated knowledge. Existing forward-only creation and refinement pipelines are insufficient for realistic deployments subject to evolving environments, policies, and privacy requirements. Treating untwinning as first-class primitives is therefore critical to maintaining consistency and trustworthiness throughout the DTN lifecycle.

\section{Resource Optimization and Semantic-Driven DTNs}

\subsection{Motivation and Basic Problems}
{
DTN synchronization consumes the resources of the links actually carrying its measurements and updates. Competition with user-access spectrum occurs when device telemetry shares access resource blocks (RBs) or a BS-to-cloud update link shares the access band. With separate wired backhaul, forwarding data already at the BS consumes backhaul and processing capacity, not user-access RBs. Thus, the resource management model must distinguish these cases before optimizing DTN synchronization and user service.
}
Meanwhile, since the wireless environment is highly dynamic due to user mobility, fluctuating traffic demands, channel fading, and network topology changes, standard wireless resource management strategies {\cite{11217298}} may not be able to balance the conflicting requirements of DTN synchronization and physical network communication quality. Therefore, dynamic spectrum resource optimization is needed for maintaining an accurate and timely DTN synchronization while ensuring high-quality communication service at physical networks. In what follows, we first introduce the state of the art on wireless resource allocation optimization for DTN. Then, the challenges and corresponding semantic-driven solutions for DTN wireless resource management are presented. Finally, an illustrative example on the design of collaborative reinforcement learning for optimizing wireless resource allocation in DTN is explained.

\subsection{State of the Art and Gaps}
{Recent works can be grouped into three categories: 1) static resource management, 2) dynamic resource management, 3) resource management with DTN feature extraction and compression, which can be
specified
} as follows.

\subsubsection{Static Spectrum Allocation for DTN and Physical Networks} Several works \cite{10375758,10949749,9491087} have optimized wireless resource allocation while considering static data transmission requirements (e.g., delay, data size) for both DTN synchronization and communication services in physical networks. In particular, the work in \cite{10375758} optimized DTN data transmission, and device transmit power to minimize the delay of device transmitting data to the DTN. The work in \cite{10949749} enhanced mobile broadband traffics for DTNs. However, these works neither consider DTN dynamics not the interference from physical networks where BSs and devices must exchange data using the same spectrum. While the authors in \cite{9491087} considered digital twin dynamics, they considered average data transmission latency thus simplifying the wireless resource optimization problem.  

\subsubsection{Dynamic Spectrum Allocation for DTN and Physical Networks} Since DTN requires frequent update to synchronize with the physical network, it is necessary to consider DTN dynamics for wireless resource management. A number of existing works \cite{10570372,ferriol2022building,zhang2023building,10619056,11309199} focused on the design of deep learning methods to model and analyze DTN dynamics (e.g., data update frequency). Meanwhile, several works {\cite{10906634,9166745,OptimizingWirelessResource2025}} designed novel reinforcement learning based adaptive wireless resource management schemes which enable network operators to change the wireless resource allocation policies according to both physical and DTN dynamics. However, most of these works considered a simplified DTN scenario such as mobile network virtualization where network dynamics caused by only user mobility.

\subsubsection{DTN Data Extraction and Compression} There are also several works \cite{11309199,10619056,10234710} that studied the use of DTN data for DTN synchronization so as to reduce the size of data collected from the physical network. In particular, the authors in {\cite{11309199}} and {\cite{10619056}} studied the use of spatial and temporal data to update DTN so as to reduce the size of data requested from the physical network. The work in \cite{10234710} focused on spatial-temporal feature fusion for IoT attack behavior identification. However, they did not consider the tradeoff between wireless resource allocation in physical network and DTN data/feature fusion and compression. Here, allocating more wireless resources (e.g., spectrum) to DTN data transmission can improve DTN synchronization accuracy. Conversely, compressing DTN data features reduces the amount of data that must be transmitted from the physical network, thereby decreasing the required wireless resource allocation.

\subsection{Challenges and Solutions}
Wireless resource management starts to play an important role in both physical and digital network domains. Although a few current works have investigated the optimization of wireless resource management for DTN, there are still several challenges (C1-C3) that must be addressed.

\subsubsection{C1: DTN Dynamic Modeling} To optimize wireless resource management for DTN, the first step is to mathematically model DTN dynamics. However, mathematically modeling the dynamics of a DTN is challenging due to the high-dimensional, nonlinear, and time-varying nature of wireless networks. The evolution of the DTN state depends on numerous coupled variables, including wireless channel conditions, user mobility, traffic demand, network topology, and resource allocation decisions, whose interactions are difficult to characterize with tractable analytical models. Furthermore, these dynamics are stochastic and partially observable, as the DTN is constructed from imperfect, delayed, and incomplete measurements collected from distributed network entities. Existing physics-based models often fail to capture the complex interactions, while purely data-driven models typically require massive amounts of training data and struggle to generalize to unseen network conditions. As a result, developing mathematical models that accurately characterize DTN dynamics while remaining amenable to wireless resource allocation optimization remains a fundamental open challenge. 

\textbf{Emerging solutions:} A promising solution is to develop a hybrid physics-informed and data-driven dynamic model for DTNs. Instead of relying solely on analytical models or purely data-driven approaches, the hybrid framework embeds wireless communication principles (e.g., channel propagation, queueing dynamics, mobility models, and network protocols) into a learning-based state transition model. The physics-based components provide interpretability, improve generalization, and reduce the amount of training data required, while the data-driven components capture complex nonlinear interactions and unknown dynamics that are difficult to model analytically. By continuously incorporating real-time observations from the physical network, the hybrid model can adapt to evolving network conditions and accurately predict future DTN states. Such a mathematical framework offers a balance between modeling accuracy, computational tractability, and adaptability, making it well suited for optimization and closed-loop control of digital network twins.

\subsubsection{C2: Physical Network Feature Extraction and Compression} Modern wireless networks generate massive amounts of heterogeneous data, including channel state information, traffic statistics, user mobility, device status, topology information, and environmental sensing data. However, transmitting all collected data to the DTN is impractical due to limited communication bandwidth, computational resources, and stringent latency requirements. Meanwhile, not all network features contribute equally to DTN synchronization accuracy, and the importance of a feature varies with network conditions, optimization objectives, and application requirements. The selected features should therefore preserve the information most critical for accurately reconstructing network dynamics while eliminating redundant or task-irrelevant information. This creates a challenging feature selection and representation problem, where the objective is to identify a compact yet informative subset of physical network features that maximizes DTN synchronization accuracy  under communication and computation constraints.

\textbf{Emerging solutions:} 
A promising direction is to develop a Semantic-Driven Twin (SDT) framework that synchronizes digital twins by transmitting network semantics rather than raw measurements \cite{zhou2026optimization}. Instead of continuously exchanging all collected network data, the physical network extracts compact semantic representations that preserve only the information essential for reconstructing the digital twin and supporting downstream optimization and control tasks. Representation learning models, such as graph neural networks, autoencoders, and transformer-based encoders, can capture the spatial, temporal, and topological dependencies among distributed network entities to learn task-relevant semantics while filtering redundant information. {Building upon these semantic representations, an adaptive semantic selection mechanism determines which semantics should be communicated according to the current network dynamics, communication resource constraints, and required twin fidelity. By jointly optimizing semantic extraction, semantic feature selection, and wireless resource allocation,
the SDT aims to reduce synchronization overhead} while preserving the information most critical to inter-realm intelligence, thereby enabling scalable, communication-efficient, and task-aware digital twin synchronization for future large-scale wireless networks.

\subsubsection{C3: Wireless Resource Management Across Hierarchical DTNs} In practice, DTNs are typically composed of multiple interconnected twins operating at different levels, such as device-level, base station-level, edge-level, and network-level twins. These hierarchical DTNs are tightly coupled since the synchronization and computation of higher-level DTNs depend on the timely updates and processing results of lower-level DTNs. Consequently, the communication, computation, and storage resources allocated to one level DTNs directly affect the performance of other level DTNs. Since all hierarchical DTNs compete for the same limited wireless spectrum, computing resources, and storage capacity, independently optimizing resource allocation at each level DTNs may lead to resource contention, synchronization bottlenecks, and suboptimal overall DTN performance. Therefore, a fundamental challenge is to coordinate resource allocation across multiple hierarchical DTN levels while accounting for their interdependencies and limited shared resources, so that the overall DTN system can maintain accurate, timely, and efficient synchronization. 

\textbf{Emerging solutions:} 
A promising direction is to empower the SDT with hierarchical reinforcement learning (HRL) \cite{11536889} for resource management in multi-level DTN systems. Rather than treating synchronization as uniform state replication, SDT represents each twin using task-oriented semantic states that capture only the information required for downstream network intelligence and control. These semantic states naturally abstract heterogeneous network measurements into a common representation while preserving their functional importance, enabling the hierarchy to distinguish critical semantic changes from redundant state updates across different DTN levels. 

Built upon this abstraction, a high-level HRL agent allocates communication, computation, and storage resources among DTN layers according to their semantic importance, synchronization urgency, and cross-level dependencies, while lower-level agents determine which semantic representations should be synchronized, at what fidelity, and at what update frequency for individual twins within each layer. Because SDT exposes semantically meaningful representations instead of raw network measurements, the HRL policy can prioritize synchronization based on the value of information rather than data volume, allocating resources to twins whose states are expected to produce the greatest improvement in network performance while postponing updates for twins experiencing only semantically insignificant changes. Furthermore, the semantic abstraction allows heterogeneous twins representing different network entities (e.g., devices, base stations, edge clouds, or network slices) to be evaluated under a unified importance metric, facilitating coordinated resource allocation across multiple hierarchical levels. {By decomposing the large-scale optimization into hierarchical decision-making tasks,
the SDT-HRL framework may reduce the action space and training complexity while naturally supporting multi-timescale resource management.} 

{
Technically, let $z=Q(E(y))$ be a quantized encoded observation, $L(z)$ its coded length in bits, and $\tilde x=D(z)$ the reconstructed twin state. A concrete design criterion is
\begin{equation}
\min_{E,Q,D,u}\;\mathbb E[J_{\rm app}(x,u)+\lambda d_W(x,\tilde x)]
 +\beta\mathbb E[L(z)]
\label{eq:semantic_rate}
\end{equation}
subject to $\mathbb E[L(z)]/\Delta\leq C_{\rm upd}$, computation limits, and a delivery deadline. Here $d_W=\|W(x-\tilde x)\|_2^2$ uses stated unit-normalizing weights, $C_{\rm upd}$ is update-link capacity in bit/s, and $\lambda,\beta$ express fidelity and bit-cost preferences. The application loss $J_{\rm app}$ can penalize robot tracking risk or missed XR deadlines. Select a feature or refinement when its expected reduction in task or fidelity loss justifies its additional bits and processing, and semantic encoding must be compared with event-triggered sampling, predictive coding, and conventional compression at matched rate and task loss.}

\subsection{Use Cases and Examples}
{
The example in~{\cite{OptimizingWirelessResource2025}} considers BSs serving mobile users while reporting partially observed network state to a cloud twin. Its shared-spectrum interpretation requires a wireless cloud-bound update link competing for the BS's RB budget. In this tutorial we explicitly adopt that assumption, but it should not be transferred to a deployment with independent wired backhaul. Device-to-BS measurement collection is a separate link. The BS selects association, RB allocation, and whether to send an update, balancing user rate against twin-state mismatch. %
}
This formulated problem is challenging to solve since each BS cannot fully observe the entire status of the physical network. To solve this problem, the authors introduced a machine learning (ML) method that integrates the {
gated recurrent units (GRUs)
} and the value-decomposition based reinforcement learning (VD-RL) method. The GRUs allow the DTN to predict its future status using historical data and to update its status when the DTN cannot receive the information of the physical network. The VD-RL method can learn from the GRU prediction accuracy to enable each BS to decide its associated users, whether to send the physical network information to the cloud server, and RB allocation, optimizing the data rate of all users in the physical network and ensuring an accurate synchronization between the physical network and the DTN. Compared to other RL methods, the designed VD-RL method allows each BS to use its partial observation, specifically, the locations of the users in its coverage, {
to coordinate decisions through a shared training objective, while global optimality is not guaranteed for nonlinear function approximation.
}

{
\begin{table}[!t]\centering
\caption{Resource tradeoff reported in~{\cite{OptimizingWirelessResource2025}}. 
}\label{tab:resource_tradeoff}
\begin{tabular}{p{0.25\columnwidth}p{0.62\columnwidth}}
\hline Quantity & Reported comparison at 12 users \\\hline
User data rate & VDN is 10.15\% higher than IQL \\
Twin mismatch & VDN error is 0.2 higher in the source scale \\
\hline
\end{tabular}
\end{table}
}
{
In this use case, the baseline is independent Q-learning (IQL), where each BS fits its own Q-function from local observations and a local reward. The value decomposition network (VDN) instead trains decomposed local Q-functions against a joint team value, providing coordinated credit assignment. The labels VDN and IQL in the table refer to these methods.
}
{
Table~\ref{tab:resource_tradeoff} retains the reported 12-user comparison but separates the rate gain from the twin-error increment. Higher user rate is obtained at a cost in mismatch, rather than a simultaneous improvement in both metrics.
}

\subsection{Summary of Learned Lessons}
This section highlights that effective resource management is essential for maintaining accurate DTN synchronization while preserving the communication performance of the physical network. Since DTN updates and user communications compete for limited wireless, computing, and storage resources, resource allocation should be dynamically adapted to both physical-network conditions and DTN synchronization requirements rather than relying on static policies. To optimize wireless resource management, we have introduced three key challenges, including DTN dynamic modeling, physical network feature extraction and compression, and wireless resource management across hierarchical DTNs. To address the first challenge, hybrid physics-informed and data-driven models can combine domain knowledge with learning capabilities to accurately characterize complex and time-varying DTN dynamics. For the second challenge, SDTs leverage task-oriented semantic representation learning to extract compact and informative network semantics, transmitting only the information necessary to preserve DTN fidelity and support downstream optimization, thereby significantly reducing synchronization overhead while maintaining synchronization accuracy. Finally, hierarchical reinforcement learning provides a promising approach to coordinating communication, computing, and storage resources across different DTN levels by jointly optimizing semantic-aware synchronization and resource allocation through interconnected hierarchical decision-making tasks. By addressing these challenges, the proposed solutions enable scalable, adaptive, semantic-aware, and resource-efficient DTN operation in dynamic wireless networks.

\section{Physical-Layer Optimization for DTNs}

The PHY layer plays a fundamentally critical role in DTNs for wireless systems, as it provides the foundational abstraction that links the physical environment with wireless signals. Accurate modeling of channel dynamics, interference, and hardware impairments at the PHY layer is essential for ensuring that the digital twin faithfully reflects real-world network behavior. 
In this section, we survey existing literature and outline the key challenges and current efforts aimed at enabling DTNs via PHY‑layer optimization.

\subsection{Motivation and Basic Problems}

Future wireless systems will increasingly support applications that tightly integrate communication, sensing, and control, including augmented and extended reality, autonomous vehicles, industrial robotics, and UAV‑based services \cite{lu2024integrated,gonzalez2024integrated,11216397}. These applications often operate in environments characterized by rapid dynamics, including mobility, blockage, interference, and fast‑varying propagation conditions that directly influence their performance. To anticipate these dynamics and adapt application behavior in real time, DTNs must incorporate PHY‑layer features that capture propagation‑environment effects with sufficient temporal resolution and fidelity.

Prior work has experimentally demonstrated that wireless signals inherently encode rich PHY‑layer information that can be leveraged to model and predict environmental dynamics. For instance, Wi‑Fi and 5G signals have been used to extract Doppler shifts and channel‑power variations for tasks such as drone detection, human activity tracking, and non-visible object recognition in indoor scenarios \cite{picazo2026802,maksymiuk2025uav,paidimarri2024eye}. Despite these advances, constructing a DTN that exploits the physical wireless infrastructure to sense such dynamics and operate in a closed‑loop manner remains a fundamental challenge.
Addressing this challenge requires DTNs to sense and update changes in the wireless propagation environment leveraging the wireless network with high accuracy and low latency. A particularly effective strategy is to leverage multi‑modal sensing embedded within the network, including video cameras, LiDAR, millimeter‑wave radars, and existing wireless communication infrastructure \cite{salehi2025darwin,yazdnian2026multi}. These sensors provide measurement data from which relevant PHY‑layer features can be extracted to construct DTNs capable of accurate prediction and control across diverse applications. Based on DT outputs, control commands are issued to sensors or actuators to adjust their operating modes, completing the closed loop, as further detailed in Sec.\ref{sec_4}. We note that this approach is consistent with the DTN vision introduced in Sec.\ref{intro}, where the objective extends beyond extracting features that merely characterize the wireless channel to identifying application‑driven, environment‑dependent PHY‑layer features that enable the creation and dynamic evolution of DTNs for a broad range of use cases beyond network modeling and prediction.

\subsection{State of the Art and Gaps}
In recent years, there has been growing interest in developing DTs that incorporate the physical (PHY)‑layer characteristics of wireless networks \cite{salehi2025darwin,li2024geo2sigmap,nokiaDT5G,ericssonDT5G,keysightDT,nvidiaAerialOmniverse,remcomWirelessInSiteDT,ansysDigitalTwinDT}. Traditional approaches based on statistical channel models, such as urban or suburban propagation models, offer low complexity but lack the accuracy required to twin a specific environment. We next review recent simulation‑based and experimental efforts addressing this challenge, as well as relevant industry-driven solutions. 

\subsubsection{PHY-layer Simulation and Experimental Efforts}
A common methodology adopted in many recent works follows a two‑step pipeline: (a) constructing a high‑fidelity 3D geometric model of the environment, and (b) applying simulation techniques over this model to estimate wireless propagation behavior.

\textit{a) Creation of geometrical models}: Recent advances have enabled the construction of realistic 3D representations of the environment using image rendering and AI/ML. For instance, GEO2SIGMAP \cite{li2024geo2sigmap} provides an ML‑based framework that integrates open‑source OpenStreetMap geographic databases with Blender, enabling efficient generation of large‑scale 3D building maps. ClickDT \cite{li2025clickdt} offers an open‑source, web‑based interface to facilitate the creation of realistic 3D scenes for DTs.
To better capture RF properties, experimental work has shown that radar and millimeter‑wave (mmWave) signals can be used to generate point clouds that naturally capture key reflectors in the environment \cite{hunt2024radcloud,guan20203d,zhao2023nerf2}. For example, \cite{guan20203d} empirically demonstrates 3D imaging with mmWave OFDM signals, while \cite{zhao2023nerf2} builds RF radiance fields from LiDAR‑generated 3D environments. Although these approaches can improve accuracy compared to purely simulation‑based 3D models, the large volume of raw sensor data significantly increases processing cost and limits agility.  For instance, one hour of point cloud data from a typical UAV‑grade LiDAR such as the Velodyne HDL-64 sensor can
result in over 100GB of data \cite{tu2019real}, which significantly limits the agility of real‑time digital‑twin updates.
To support robot mobility in dynamic environments, DARWIN \cite{salehi2025darwin} enables real‑time DT generation by having robots transmit local sensor information to the autonomous edge and receive motion commands. Despite their higher accuracy and ability to support dynamic environments, relying on dedicated sensors can be costly in terms of deployment and spectrum usage.

\textit{b) Link-level simulation tools}: To estimate PHY‑layer features over the generated 3D model, ray tracing (RT) has been widely used. RT is a physics‑based method that simulates how waves travel and interact with surfaces by tracing their paths through a 3D environment \cite{li2024geo2sigmap}. A well‑known RT tool is Sionna, a GPU‑accelerated open‑source library for link‑level simulations based on TensorFlow that integrates differentiable RT for radio‑wave propagation \cite{hoydis2023sionna}.
Despite their popularity, RT‑based methods suffer from poor trade‑offs between computational complexity and accuracy in complex environments \cite{li2024geo2sigmap}. Although sensor data can be incorporated into such models to boost accuracy, it often comes from disaggregated sources, such as LiDAR databases, and is not updated in real time.
To address this, and building on the concept of RT, a family of works has been inspired by the success of neural networks in describing optical fields in computer vision. For instance, \cite{zhao2023nerf2} proposes the use of neural radio‑frequency radiance fields to estimate RF propagation in static volumetric regions constructed from 3D LiDAR measurements, whereas \cite{pumarola2021d} extends this concept to dynamic scenes and demonstrates it in simulation. However, the high data requirements and slow training and update times of these approaches limit their applicability for the creation of DTNs. Thus, efficiently incorporating channel dynamics into such 3D geometrical models remains an open challenge.\cite{li2024geo2sigmap}.

{
\textit{c) Multimodal reconstruction and online calibration}:
The TUM--Oulu--ASU experimental study by Andrei et al.~\cite{Andrei2025Multimodal} combines RGB-D observations and visual-SLAM poses with semantic segmentation and database-based material assignment. Nimbus traces propagation directly through the labeled point cloud. Indoor SDR measurements validate a 5G robot-link and monostatic sensing setup using EVM and ranging error. This is evidence for a concrete reconstruction and validation pipeline, not proof of unrestricted city-scale online operation.
More generally, synchronized RGB-D or LiDAR observations can update surfaces and object trajectories, and radio residuals can calibrate electromagnetic parameters that visual labels alone cannot identify reliably. A representation-update cycle registers observations in a common frame, detects structural changes, updates geometry and material hypotheses, predicts channel responses, and checks residuals on held-out measurements. A changed object pose is a state update when the object already exist, and a newly discovered reflector or revised material law changes the representation. Occlusion, SLAM drift, ambiguous material labels, and hardware impairments must be tracked as uncertainty. A candidate update is accepted only after fidelity and computation checks, with rollback if it degrades predictions. Environment-aware twins have also been proposed for handover, edge migration, and video preloading~\cite{Alkhateeb2023RealTime}, illustrating how reconstruction supports decisions beyond PHY parameter tuning.
}

\subsubsection{Industry-Driven Efforts}

Several industry initiatives have initiated the exploration of DTs for wireless propagation and PHY‑layer modeling. Telecommunications vendors such as Nokia and Ericsson have introduced early DT platforms for 5G that rely on network telemetry, drive‑test data, and propagation models to estimate coverage, interference, and radio‑link performance \cite{nokiaDT5G,ericssonDT5G}. In parallel, measurement‑driven DT solutions from companies such as Keysight provide network‑level digital twins that integrate RF measurements, channel modeling, and scenario‑based simulation for performance prediction and optimization \cite{keysightDT}. However, these systems primarily support network planning and optimization at a slow scale rather than real‑time PHY‑layer prediction.
Notable efforts focus on high‑fidelity radio‑environment simulation. NVIDIA Sionna \cite{hoydis2023sionna} and Aerial Omniverse \cite{nvidiaAerialOmniverse} provide GPU‑accelerated, differentiable ray‑tracing engines capable of modeling multipath propagation, blockage, and mobility effects for mmWave and sub‑THz systems. Commercial ray‑tracing and EM‑simulation tools such as Remcom Wireless InSite \cite{remcomWirelessInSiteDT} and Ansys Digital Twin \cite{ansysDigitalTwinDT} are widely used for site‑specific channel modeling and scenario‑based evaluation. {These tools provide propagation-modeling components, while online sensing, calibration, and closed-loop orchestration depend on their surrounding implementation and should not be inferred from ray tracing alone.}

\subsection{Challenges and Emerging Solutions}
{
In the six-stage lifecycle of Fig.~\ref{fig:DTN_design}, PHY sensing supplies stage 1, feature alignment supports stage 2, and geometry and propagation reconstruction form stage 3. These are recurring stages, not a construction-first pipeline. Their output supports simulation, decisions, and actuation in stages 4--6. The following challenges (C1-C3) address both rapid state tracking and slower changes in the physical representation.
}

\subsubsection{C1: Efficient Extraction of Wireless Telemetry Data}
A critical challenge for the creation of DTNs is how to \textcolor{black}{dynamically} leverage wireless network resources without significantly increasing the network complexity, cost, power consumption, and spectrum usage. Conventional sensing platforms, such as drone-based network, mobile radars, or other dedicated wireless measurement units, provide only partial coverage and incur substantial operational and \textcolor{black}{deployment} cost, making them impractical for \textcolor{black}{agile}, large‑scale, real‑time DTN operation. Moreover, power efficiency is a key design constraint in next‑generation wireless systems, raising concerns about relying on energy‑intensive radar or drone‑based sensing for wide‑area operation. Additionally, the limited available spectrum must be considered when leveraging wireless systems that share bandwidth with communication signals. These factors become a roadblock to efficiently extracting network data from conventional sensing platforms to support DTNs for applications in dynamic wireless environments. 

\textbf{Emerging solutions:} Integrated Sensing and Communication (ISAC). ISAC \textcolor{black}{is an emerging solution} for next‑generation wireless systems and has been included in 6G studies by 3GPP \cite{eren2025integrated}. In ISAC, communication waveforms, infrastructure, and signal processing are leveraged to perform additional sensing tasks. Therefore, ISAC can be resource‑efficient alternative to conventional sensing platforms, as it inherently aligns sensing operations with communication procedures. This allows the network to extract environmental information without deploying dedicated sensing hardware, thereby reducing overhead and enabling persistent, low‑cost monitoring across the entire \textcolor{black}{wireless environment}. A relevant example is the use of 5G networks to perform additional sensing tasks such as concealed‑object detection \cite{paidimarri2024eye}. Specifically, in the 5G mmWave 5G band (FR2), the network periodically triggers a standard beam‑scanning procedure every 20–160 ms. In this procedure, a 5G transmitter broadcasts directional signals to announce itself in the network. These signals naturally capture channel dynamics within the channel coherence time across multiple spatial directions. A 5G‑enabled receiver can then measure the received signals and extract communication features such as the channel impulse response.  
These features can be used to perform sensing tasks without the overhead of a dedicated sensing system. 

\begin{figure}[]
    \centering
    \includegraphics[width=\columnwidth]{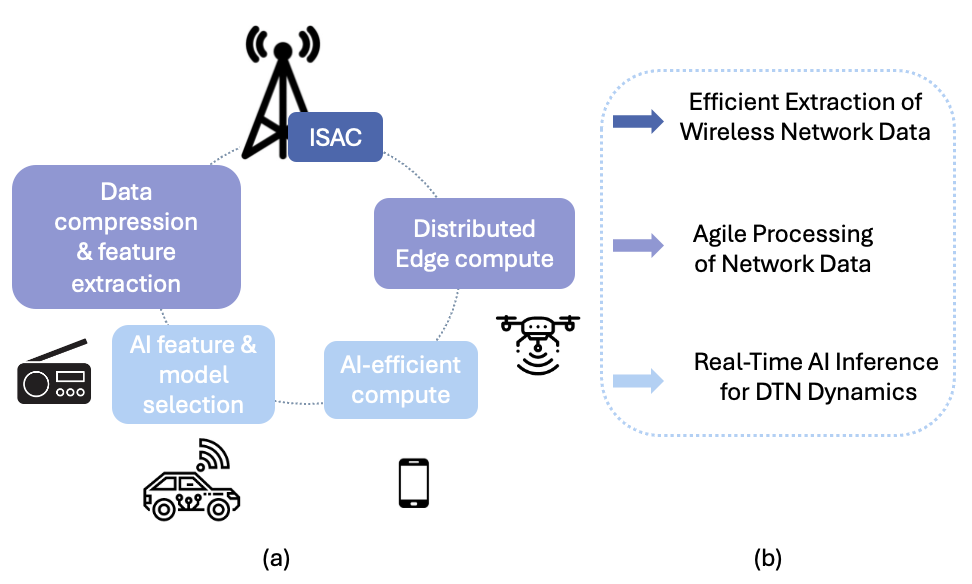}
    \caption{\small Overview of (a) emerging PHY-layer solutions and (b) their corresponding challenges toward enabling dynamic Digital Twin Networks.}
\label{fig:phylayer}
\end{figure}

As one example, Panoptic \cite{guo2025panoptic} experimentally demonstrates the acquisition of unique spatial information through pseudorandom sidelobe perturbations while  maintaining a directional mainlobe toward the intended communication nodes leveraging a technique called as compressive sidelobe forming. Moreover, in \cite{gallyas2024angle} the authors estimate the Angle-of-Arrival (AoA) of a received signal by rapidly creating a succession of beam-patterns within the duration of a single OFDM symbol in 5G-compliant waveforms.

\subsubsection{C2: Agile Processing of Network Data}
Another key challenge is the agile extraction of network data, which requires performing (a) efficient and application‑relevant sensor data compression and feature extraction, and (b) provisioning the distributed compute infrastructure needed to keep pace with the dynamics of DTNs at the edge. 

\begin{enumerate}[label=\alph*), leftmargin=*]

    \item Raw IQ samples or full‑resolution sensor outputs are too large to support timely DTN updates, making compact, task‑oriented representations essential. Processing raw sensor data, such as LiDAR point clouds, radar sweeps, or mmWave channel scans, is computationally demanding and generally incompatible with the latency requirements of real‑time DTN operation. To remain agile, the system must extract a compressed set of features that are directly relevant to the application.
    
    \textbf{Emerging solutions:} Sensor Data Compression and Feature Extraction.
    
    A promising direction is feature compression where information is distilled either post‑collection through lightweight neural encoders or directly at the sensor via embedded feature‑extraction modules. For instance, in the latter case, the authors \cite{xu2026multiply} only measure the compressed representation of the feature space using dedicated hardware. This significant reduces the amount of data that passes through the Analog-to-Digital converter, that is the most power hungry component of the data processing. 
    Moreover, several major research directions focus on enhancing concise feature representations. These include attention techniques, distillation for data-efficient training of vision transformers, and multi-scale convolutional neural networks (CNNs). For instance, the work in \cite{chen2021crossvit} presents CrossViT, a dual-branch vision transformer for learning multi-scale features, to improve the recognition accuracy for image classification.

    \item Feature extraction and compression must occur as close to the sensor as possible to maintain DTN agility \cite{kim2023real}. Processing raw data streams, such as LiDAR point clouds, radar sweeps, or mmWave channel scans, at centralized locations introduces latency, increases backhaul load, and requires transmitting large volumes of information to upper layers of the DTN. Without local processing, the DT cannot remain synchronized with the physical environment, especially under real‑time constraints. 

    \textbf{Emerging solutions:} Distributed Edge Sensor Processing.

    Deploying compute resources at the network edge enables real‑time execution of lightweight signal‑processing pipelines, neural encoders, and task‑oriented feature‑compression modules. By extracting compact, application‑relevant representations directly at or near the sensor \cite{salehi2025darwin}, only essential information is forwarded to the DTN. This reduces communication overhead, alleviates backhaul load, and helps keep the DTN aligned with the physical system without exceeding network or compute constraints.
\end{enumerate}

\subsubsection{C3: Real-Time AI Inference for DTN Dynamics}

{
Inference deadlines depend on the task and prediction horizon. For instantaneous channel tracking or beam selection, channel coherence can constrain the useful age of the estimate. Geometry and blockage tracking depend on object motion and sensing latency; queue scheduling depends on packet deadlines; traffic forecasts and RAN sleep decisions operate over their own aggregation and switching intervals. Security decisions depend on monitoring windows and the tolerated attack-response delay. The total sensing, transport, inference, decision, and actuation delay must fit the task's freshness or control budget, thus channel coherence is not a universal DTN deadline.
}

\begin{enumerate}[label=\alph*)]
    \item \textbf{Emerging solutions:} \textit{AI \textcolor{black}{Feature Selection.}}
    
    Feature selection is essential for maintaining both interpretability and computational tractability. Pruning redundant or low‑information features reduces model dimensionality, accelerates inference, and mitigates overfitting without degrading predictive performance. Task‑aware model adaptation further refines this process by selecting only the features relevant to the current operational context—such as mobility regime, propagation conditions, or service‑level objectives—allowing the DT to dynamically tailor its inference pipeline to the underlying network state \cite{alhamam2025twinguard}.

    \item \textcolor{black}{\textbf{Emerging solutions:} \textit{AI Model Selection}.}
    
\textcolor{black}{Modern DT frameworks increasingly adopt adaptive model selection, choosing among multiple candidate models based on their expected accuracy, complexity, and relevance to the current network state. For instance, recent work \cite{chen2023digital} demonstrates that different models exhibit varying sensitivity to feature availability, measurement noise, and mobility‑induced distortions. By estimating feature‑level accuracy and linking it to each model’s performance profile, the DT can dynamically select the predictor that best matches the instantaneous feature quality at the edge. This adaptive selection maintains inference accuracy under imperfect observability while reducing computational overhead by avoiding unnecessarily complex models when simpler ones suffice. 
    }

\item \textbf{Emerging solutions:} \textit{AI Computational efficiency}.

Real‑time inference under sub‑millisecond latency budgets requires hardware‑accelerated execution. GPUs, FPGAs, and emerging AI accelerators provide the parallelism and throughput necessary to sustain continuous prediction at a high rate \cite{samanta2024survey}, as imposed by fast‑fading channels and ISAC‑enabled sensing updates. When combined with model‑level optimizations, these platforms enable DTNs to preserve predictive fidelity while operating within the stringent temporal constraints of next‑generation wireless systems.

\end{enumerate}

\subsection{Summary of Lessons Learned}

The potential development of real‑time DTNs reveals several fundamental challenges about how future wireless systems must sense, interpret, and react to rapidly evolving environments. First, PHY‑layer awareness is indispensable. Modern wireless applications, ranging from XR to autonomous mobility, operate in environments characterized by fast dynamics, where mobility, blockage, and interference continuously reshape the propagation landscape. DTNs must therefore incorporate PHY‑layer features that capture these variations with sufficient temporal fidelity. Prior work demonstrates that wireless signals inherently encode rich environmental information, but leveraging this information in a closed‑loop DTN remains challenging.

Second, efficient data extraction is essential for scalable DTN operation. Raw IQ samples, LiDAR point clouds, radar sweeps, and mmWave scans are too large and computationally demanding to support timely DTN updates. This motivates the need for compact, task‑oriented feature representations that preserve only the information relevant to prediction and control. Emerging solutions such as ISAC enable the reuse of communication waveforms for sensing, reducing hardware overhead and enabling persistent, low‑cost environmental monitoring. Complementary advances in feature compression, ranging from lightweight neural encoders to hardware‑level compressed sensing, are promising techniques to further reduce data volume and power consumption.

Third, feature extraction must occur as close to the sensor as possible. Centralized processing introduces latency and backhaul congestion that break DTN synchronization. Distributed edge processing allows sensors and nearby compute nodes to execute lightweight signal‑processing pipelines and neural encoders in real time, forwarding only compact, application‑relevant features. Considering and evaluating this architectural shift is critical for maintaining DTN agility under tight latency constraints.

{
Fourth, computational efficiency must be assessed against a task-specific end-to-end deadline. Channel-coherence limits can apply to rapid PHY tracking, whereas motion prediction, queue control, energy management, and security use different horizons. Feature and AI model selection should therefore report both prediction quality and total update-to-action latency, including transport and actuation, rather than applying a single coherence-time bound to all DTN operations.
}

Finally, hardware acceleration is indispensable for sustaining sub‑millisecond inference. GPUs, FPGAs, and emerging AI accelerators provide the parallelism required to support continuous prediction at rates dictated by fast‑fading channels and ISAC‑enabled sensing updates. When combined with model‑level optimizations, these platforms can become key enablers for DTNs to maintain predictive fidelity while operating within the stringent temporal, power, and spectrum constraints of next‑generation wireless systems.

\section{DTN-Guided Network Management and Feedback Optimization}
\label{sec_4}

\subsection{Motivation and Basic Problems}
The evolution toward 5G and future 6G wireless networks has significantly increased the complexity of network operation due to massive device connectivity, heterogeneous services, and highly diverse Quality of Service (QoS) requirements. In this regard, next-generation wireless infrastructures must simultaneously support latency-sensitive applications, high-throughput services, and large-scale IoT deployments. However, traditional network management paradigms mainly operate based on reactive control loops and extensive human intervention \cite{traffic_eng_survey,native_design}. 

The central problem in DTN-guided network management is related with the role of feedback. In traditional paradigms, telemetry is typically consumed as a passive monitoring output for reporting and alarms, rather than as an active control variable that directly guides closed-loop optimization. Despite the integration of AI/ML for prediction and automation, the lack of systematic feedback policies such as when and what to update, and how to react under changing circumstances still lead to unstable control actions. This challenge even become more important for heterogeneous and dynamic service environments, making static optimisation policies insufficient.

At this point, it is obvious that traditional network management paradigms are reaching their limits by creating a critical need for shifting toward DTN-guided network management. Here, DTN-guided network management provides proactive and feedback-aware network control. More specifically, DTNs provide a closed loop optimization via the feedback link, and as a result of this, timely and accurate actions are provided to the physical network. With this, the management functions \textit{fault}, \textit{configuration}, \textit{accounting}, \textit{performance}, and \textit{security} (FCAPS) can be evaluated and optimized in an online manner. For instance, from FCAPS perspective, the DT guidance can provide predictive fault detection, enhanced configuration through what-if capability, context-aware accounting, proactive performance management, and safe security validation. Therefore, we highlight the role of DTN in the network management and feedback optimisation in this section.

\subsection{State of the Art and Gaps}
In this section, we provide a set of prior studies of DTNs from an intelligent closed-loop operation perspective, focusing on the barriers to providing full synchronisation and effective feedback optimisation.

DTN for network management represents a digital embodiment of the entire network architecture, constructed using representations derived from data collected through sensors and telemetry tools. This data includes information on user devices, real-time network states, traffic patterns, data rates, and operational logs, which collectively enable the accurate modelling of the physical network within its virtual counterpart. As a result, DTN serves as an effective tool for network monitoring, control, and optimization through emulation and simulation. A fundamental step in realizing an effective DTN for network management is model selection and architectural design, as it directly impacts scalability, accuracy, and real-time operability. Studies such as \cite{ModelSelectionOffloading2025} focus on flexible deployment of DTN models by introducing model splitting mechanisms that enable real-time execution across distributed environments. Similarly, machine learning–driven modeling approaches, including federated learning and Multi-Layer Perceptron (MLP)-based DTN construction, have been explored in \cite{InternetFederatedDigital2024,GenerativeAIDrivenDigital2024,10143356} to address data heterogeneity under energy-awareness constraint.

Beyond model selection, recent works emphasize native DTN architectures for future 6G networks. In particular, \cite{DesignAIdrivenNetwork2024} highlights the role of AI in defining the fundamental DTN structure by enabling secure closed-loop operation and incorporating proactive defense mechanisms against network threats. Building upon these concepts, the authors of {\cite{native_design}} make several key contributions by proposing a 6G-native DTN logical framework and functional architecture. Their work introduces four DTN functional node (DTNF) designs, a complete operational procedure, and three deployment modes, establishing a systematic foundation for achieving high-level network autonomy in 6G systems. Furthermore, the paper identifies high-performance twin modeling and intelligent native orchestration and management as critical enabling technologies, and provides preliminary innovative solutions in these areas. By discussing emerging research directions and novel design paradigms, this work offers a comprehensive reference framework that guides future research on autonomous DTN-enabled network management.

For effective operation, a DTN must remain sufficiently synchronized with the physical network through updates from base stations. However, frequent synchronization incurs non-negligible communication and computational overhead, leading to resource contention between user traffic and DTN update processes. To address this, several studies have proposed optimization-based synchronization mechanisms. For example, in {\cite{OptimizingWirelessResource2025}}, spectrum allocation is jointly optimized to regulate DTN synchronization while preserving model accuracy. In such frameworks, feedback from the digital twin to the physical network enables adaptive update signaling and closed-loop control. In this regard, learning-based approaches have further been explored to support synchronization decision-making. In particular, gated recurrent units (GRUs) have been employed to determine when base stations should transmit state updates to the digital twin, based on temporal network dynamics and resource availability. Complementary to these approaches, standardized data representation models such as YANG have been adopted to facilitate structured, interoperable information exchange between the physical network and its digital counterpart \cite{BuildingDigitalTwin2025,t6conf}. Together, these works demonstrate the importance of optimized synchronization and feedback mechanisms in DTN-based network management, while laying the groundwork for intelligent closed-loop operation.

\subsection{Challenges and Solutions}
\subsubsection{Key Challenges}
\paragraph{C1. Synchronization Timing and Communication Overhead Dilemma} DTNs rely on data from the physical network to construct and maintain a virtual representation of network states and dynamics. This representation is typically learned from historical network data, allowing the digital twin to capture long-term behavior and patterns of the physical network. However, due to the highly dynamic nature of wireless environments, historical data alone is insufficient to accurately reflect instantaneous network conditions, particularly under time-varying traffic loads and channel variations.

To preserve synchronization fidelity, the digital twin must be periodically updated with recent measurements from base stations through a robust feedback mechanism. Although frequent updates improve the accuracy of the digital twin, they introduce additional communication overhead and compete with user traffic for limited network resources. In contrast, infrequent updates reduce overhead, but may result in model drift, leading to inaccurate predictions and suboptimal network management decisions. Consequently, a fundamental challenge lies in determining when and how base stations should transmit state updates, as well as in establishing an efficient two-way communication framework between the physical network and its digital counterpart that balances synchronization accuracy and feedback cost.

\paragraph{C2. QoS Heterogeneity}Modern wireless networks support a mix of services with very different QoS needs, such as low latency and high reliability for real-time applications, and high throughput for data-intensive services. In addition, user devices differ in capability, mobility, and traffic behavior, network resources remain limited, and vary over time. These differences lead to conflicting QoS requirements, where prioritizing one service can directly degrade the performance of others sharing the same resources. In particular, latency- and reliability-critical services require quicker feedback and more frequent control adjustments, whereas delay-tolerant services can operate with looser control. This imbalance complicates DTN-guided network management and feedback optimization, as a single optimization policy or decision model may fail to capture the diverse adaptation needs of all services. As a result of this, ensuring consistent QoS across heterogeneous services under dynamic network conditions remains a significant challenge.

\paragraph{C3. Lack of Model Evaluation Mechanisms} An effective digital twin  for network management requires not only accurate modelling but also continuous performance evaluation and control of representation fidelity. As DTN-based management increasingly relies on predicted network states to reduce feedback overhead, inaccuracies in the digital twin model can lead to suboptimal or unsafe control decisions. Therefore, measuring the deviation between the digital twin and the physical network becomes a critical challenge, particularly in feedback-optimized systems where real-time updates are limited. In these circumstances, without proper evaluation metrics and accuracy control mechanisms, the digital twin cannot reliably determine when additional feedback is required, undermining both synchronization efficiency and QoS assurance.

\subsubsection{Solution I: Learning-Assisted Proactive Synchronization}
To overcome the synchronization timing and communication overhead dilemma, the integration of AI algorithms can provide intelligent decisions to guide the synchronization timing while also considering the feedback cost. In this regard, machine learning–based prediction mechanisms can be employed to reduce the frequency of state updates from the physical network. For example, Long Short-Term Memory (LSTM) networks learn temporal patterns in network state information and predict the evolution of key system parameters. Based on these predictions, the digital twin can determine when the base stations need to transmit updated physical network state information, instead of relying on continuous feedback. With this, the digital twin can provide proactive actions for the synchronization timing. As a result, this proactive update strategy significantly reduce communication overhead while maintaining sufficient synchronization accuracy between the physical network and its digital twin.

\subsubsection{Solution II: Service-Aware Multi-Objective DTN Optimization} To address QoS heterogeneity problem, multi-objective optimization approaches have high potential to balance diverse service requirements. For example, services are organized into a priority hierarchy, with URLLC assigned the highest priority due to its strict latency and reliability constraints, followed by eMBB and mMTC \cite{QoSreq}. High-level optimization enforces service-aware QoS constraints, while lower-level optimization allocates remaining resources to optimize competing objectives such as throughput and energy efficiency. As a result, the problem can be formulated as a multi-objective optimization problem to balance trade-offs among heterogeneous services. 

\subsubsection{Solution III: Metric-Driven Model Evaluation and Control} To address model evaluation and accuracy control in DTN-based network management, metric-driven evaluation frameworks should be adopted to quantify the fidelity and timeliness of the digital twin. For instance, a set of well-defined digital twin performance metrics should be chosen from the current digital twin literature by considering the use case  \cite{dtaas_survey}. Here, the key metrics can include \textit{twin fidelity}, which measures the accuracy of the digital twin with respect to the physical network; the \textit{twinning rate}, representing the frequency of updates from the physical network, and \textit{age-of-twin}, defined as the time elapsed since the last physical update. By controlling these metrics, the DTN can simulate different feedback strategies and identify operating points that balance representation accuracy and communication overhead.

\subsection{Use Cases and Examples}
\subsubsection{DT-enabled {
Radio Access
} Network (RAN) Management} RAN management use case involves monitoring and optimisation of network components, including base stations and antennas, to guarantee user availability in accordance with their demand. In this use case, the DT is treated as a service that creates a virtual representation of the RAN by continuous mirroring of the behaviour and patterns of base stations, including user load, on/off status, energy consumption, and throughput. As part of a broader DTN, this virtual replica enables real-time monitoring and informed decision-making for RAN control.

\begin{figure}[htbp!]
    \centering
    \includegraphics[width=\columnwidth]{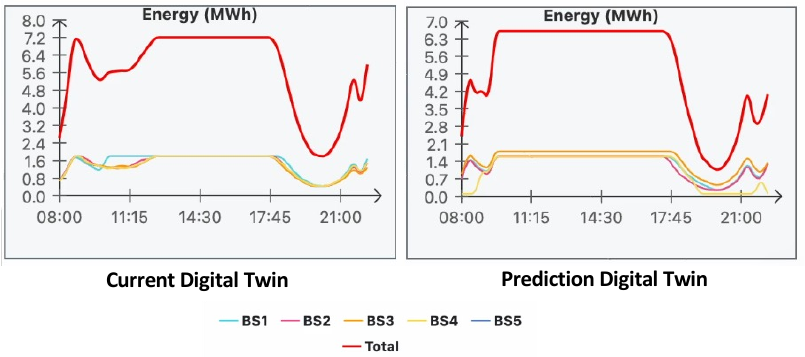}
    \caption{{\small Illustrative RAN energy monitoring and prediction trace.}} 
\label{fig:ran_dt}
\end{figure}

{
The energy trace in Fig.~\ref{fig:ran_dt} illustrates monitoring and prediction. Its right-hand panel is labeled as a prediction, but the available trace does not identify an executed action, a controlled counterfactual, or an energy-saving baseline. It therefore does not infer measured energy savings or maintained service guarantees from this panel. The dimensionless settings are not used as measured update frequency or fidelity because their units, normalization, and estimation procedures are unspecified.
A realizable energy-control design would collect BS load, queue and power counters, update the twin, forecast demand, evaluate cell-sleep or carrier-activation candidates, and execute a selected configuration only if coverage, capacity, and delay constraints are satisfied. Energy reduction then comes from the changed cell or carrier state, not merely from adjusting a twinning metric. 
}

\subsubsection{DT-enabled Intelligent DDoS Detection Mechanism for Autonomous Core Networks} In this use case, a digital twin can be employed to support the detection and mitigation of Distributed Denial-of-Service (DDoS) attacks in a networked system with interconnected computers, routers, and organisational gateways. Usually, physical networks continuously receive data traffic with varying volumes and patterns, making it challenging to distinguish between benign surges and malicious activity in real time. Here, DT acts as a virtual representation of the network, mirroring traffic statistics, flow characteristics, and routing behaviour to enable continuous monitoring and analysis of network conditions.

\begin{figure}[htbp!]
    \centering
    \includegraphics[width=\columnwidth]{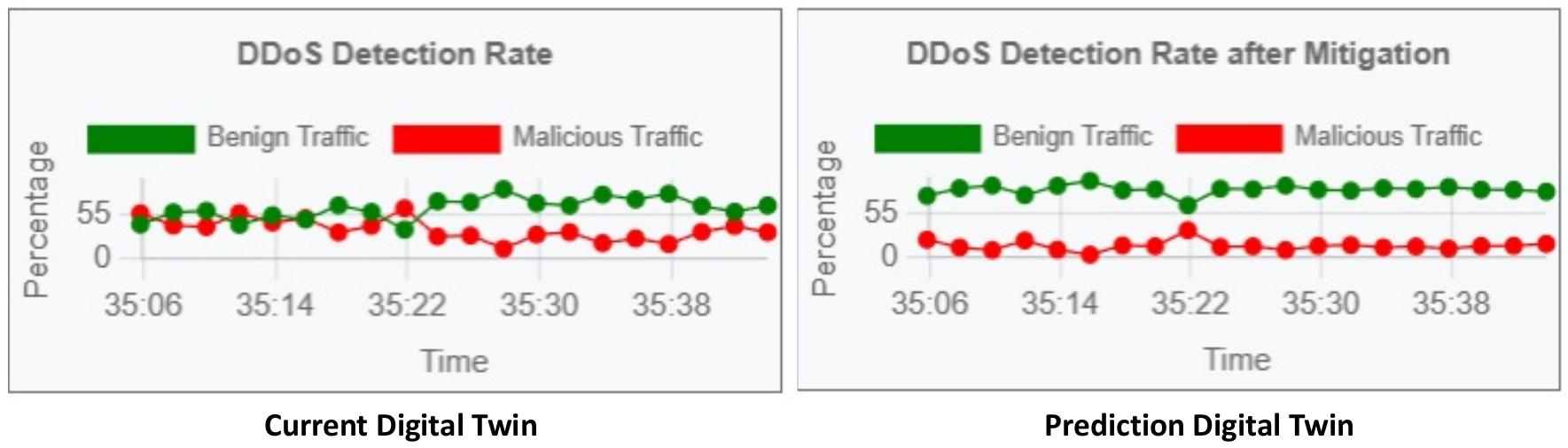}
    \caption{{\small Illustrative observed and predicted traffic shares. 
    }}
\label{fig:ddos_dt}
\end{figure}

{
Figure~\ref{fig:ddos_dt} shows a use case of traffic composition and a predicted composition. A malicious-traffic share is a fraction of traffic in a specified window; it is not detection precision, recall, false-positive rate, or response latency.
An closed-loop implementation collects flow counters, updates a traffic model, computes an anomaly score, and asks a policy to install a rate limit, filter, or route diversion at an enforcement point. Subsequent admitted traffic and benign-service loss provide feedback. A z-score threshold is only a possible detector and must be evaluated under nonstationary benign load. Note that validation requires labeled attacks, an explicit mitigation action, detection and false-alarm counts, time-to-enforcement, and benign traffic loss. More DTN-assisted management examples can be found in Fig.~\ref{fig:feedback_path}.
}

{
\begin{figure*}[t]
    \centering
    \includegraphics[scale=0.25]{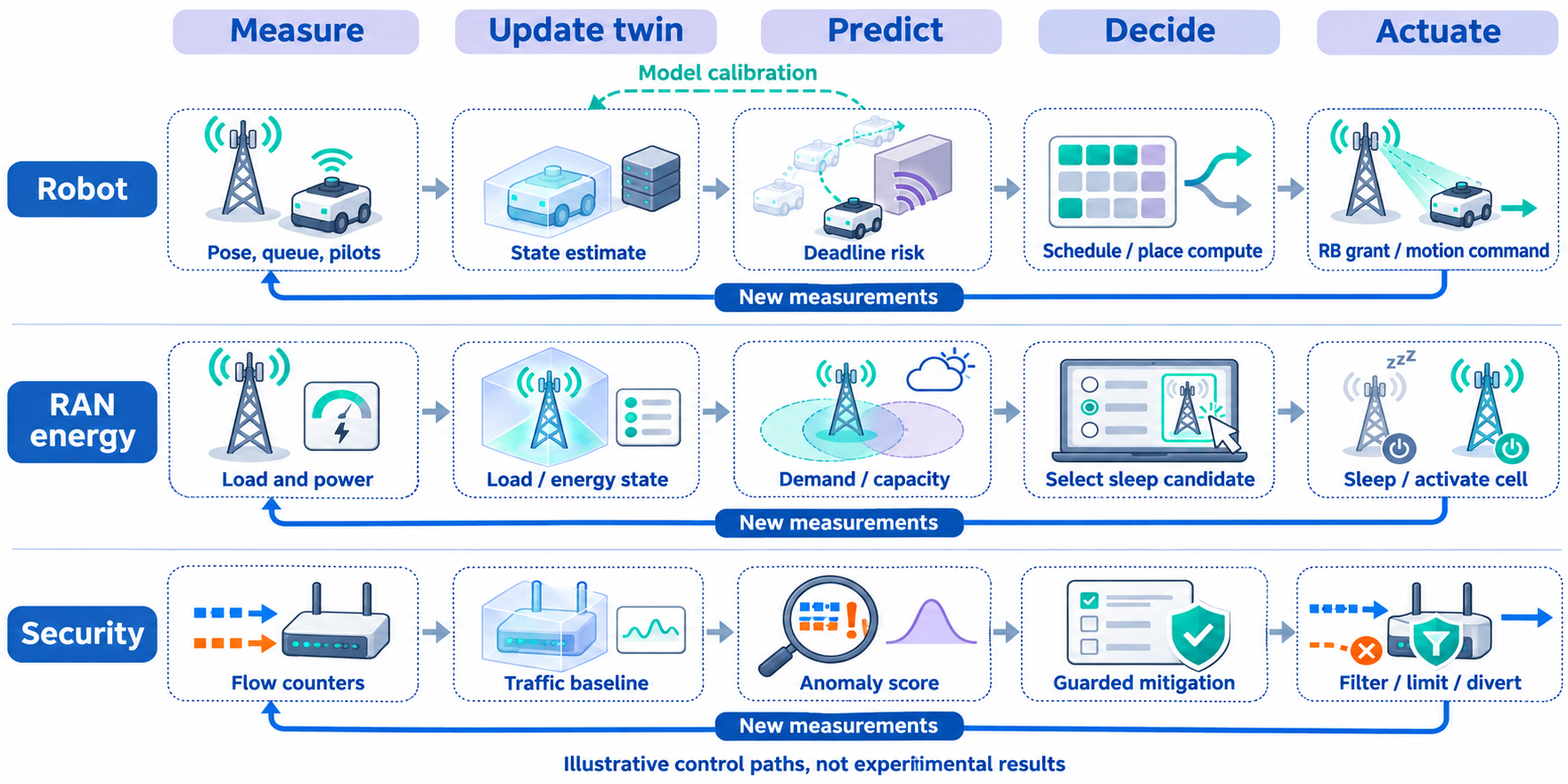}
    \caption{Illustrative closed-loop control paths for robot service, RAN energy management, and security. Each path links measurements, twin updates, predictions, decisions, and physical actions, where action outcomes then provide new measurements. Dashed feedback denotes model calibration.}
    \label{fig:feedback_path}
\end{figure*}
}

\subsection{Summary of Lessons Learned}
{Overall, there are several key lessons learned regarding DTN-guided network management and feedback optimization. First of all, DTNs become significant only when the feedback loop to the real environment is treated as an active control mechanism. More specifically, this feedback mechanism is required to perform real-time operational actions which is the main feature of a DT that
distinguishes it from traditional monitoring tools.} From a practical perspective, the ability \textit{when} to synchronize and \textit{how} to translate the insights produced by the DT into real-world system actions serve a proactive closed-loop optimization within the network management. Furthermore, the synchronization in DT should not be considered individually, rather it should always be considered with the system resources and QoS heterogeneity. Regarding this, frequent updates improve fidelity but compete with user traffic and compute budgets, while infrequent updates lower system accuracy and degrade decision quality.  Therefore, learning-based and service-aware policies are essential to adapt feedback timing and control granularity.  Instead of relying on uniform update schedules, proactive synchronization methods should be discovered by also considering multi-objective optimization approaches. Finally, reliable DTN operation requires metric-driven model evaluation and control. In this regard, the DTN must quantify both representation fidelity and timeliness using well-defined performance indicators such as \textit{twin fidelity} and \textit{twinning rate}. These metrics enable the twin to evaluate how much it reflects the physical network, to compare alternative feedback strategies through what-if analysis, and to select operating points that balance decision accuracy against communication overhead. Without such metric-based evaluation, DTN-guided control may rely on inaccurate states and thus produce suboptimal management actions.

\section{Standardization and Enabling Tools}
\label{sec7}
\subsection{Motivation and Basic Problems}

A natural bridge between the conceptual notion of DTNs and concrete implementations is emerging in the form of standardization frameworks and enabling tools. On the one hand, International Telecommunication Union (ITU) has defined DTN architecture, capability levels, and 5G integration, while the 3rd Generation Partnership Project (3GPP) embeds DTNs into the 5G management system. In parallel, 3GPP's channel modeling work highlights both the importance and the current absence of standardized, physics-based ray-tracing models, which are essential for site-specific wireless digital twins. On the other hand, practical tool chains demonstrate how these standards can be instantiated in practice. 

Beyond modeling fidelity as introduced previous sections, DTNs must integrate tightly with existing network management and orchestration frameworks. This integration requires standardized interfaces for telemetry collection, configuration control, and analytics feedback, as well as well-defined workflows that embed the digital twin into operational loops. Without standardization, digital twins risk becoming isolated simulation tools that cannot scale across vendors, domains, or deployment scenarios. These challenges motivate the need for standardized DTN architectures, capability definitions, and management interfaces, which have become the focus of recent efforts within ITU-T and 3GPP. In this section, we introduce the key DTN standardization works and challenges, and then connects them to representative tools that realize DTNs in research and product settings.

\subsection{State of the Art and Gaps}

\subsubsection{\underline{ITU Standardization of DTNs}}

ITU telecommunication standardization sector (ITU-T) has developed a set of recommendations on DTNs. In the context of wireless digital twins, these recommendations provide the standards-based framework within which concrete radio access network (RAN) and end-to-end mobile network twins can be designed, deployed, and compared. In the following, we introduce the ITU-T work that defines this DTN framework.

\paragraph{Architecture}

A formal architectural treatment of network digital twins has been developed in the ITU-T Recommendation Y.3090 {\cite{ITUT2022Y3090}}. It introduces the notion of DTN and specifies its reference architecture and requirements. A DTN is defined as a virtual counterpart of a physical network that is continuously driven by data, models, and standardized interfaces so as to maintain a near real-time, bi-directional correspondence between the physical network and its twin. This twin is not only used for offline analysis and simulation, but is also expected to support diagnosis, emulation, and control of the physical network. In the wireless context, the physical network may be a single-domain RAN, a multi-domain end-to-end 5G system, or even a specific network slice, while the DTN models the corresponding RAN, core, and transport abstractions.

\begin{figure}[htbp!]
    \centering
    \includegraphics[scale=0.15]{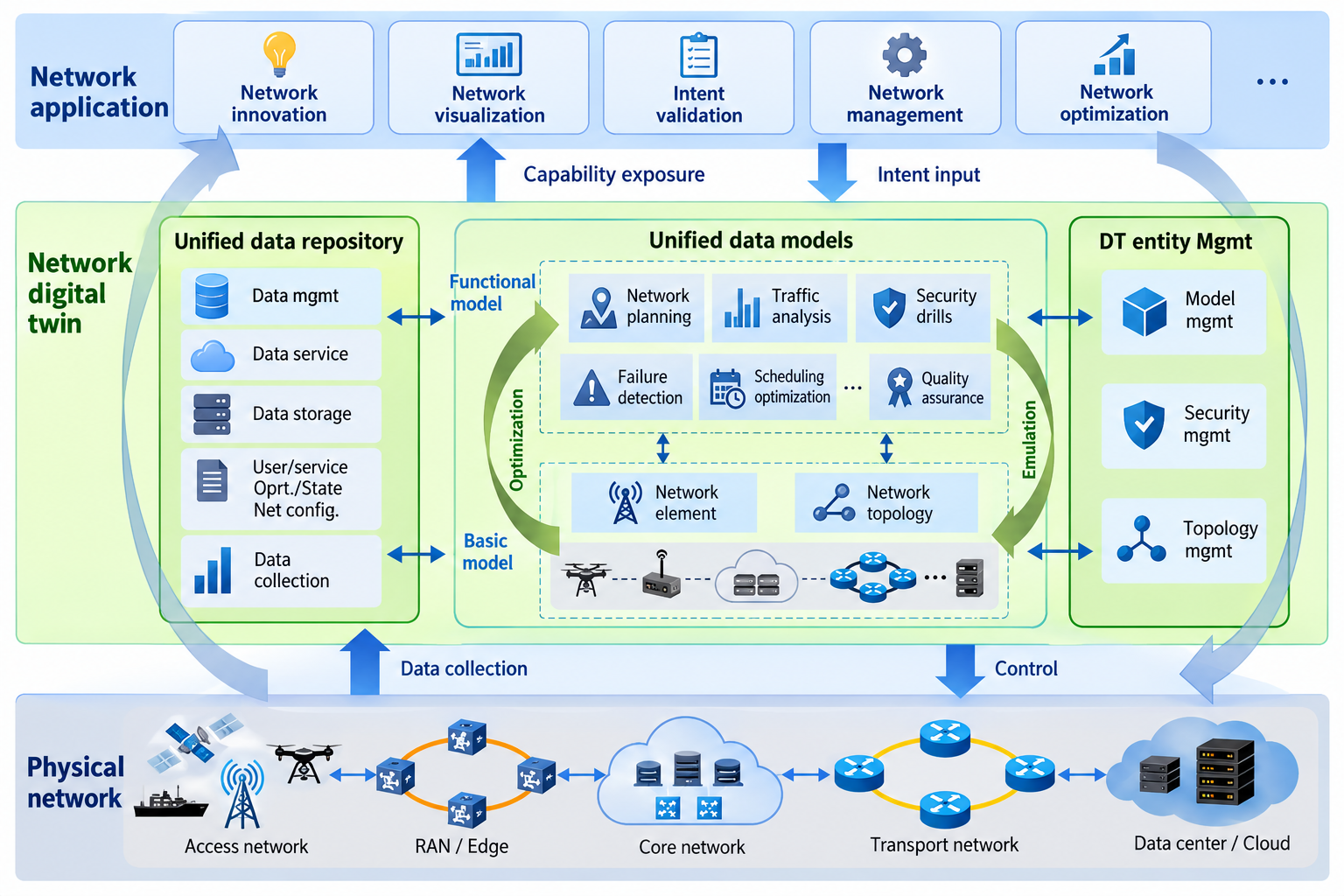}
    \caption{\small A ``three-layer, three-domain, double closed-loop structure for DTN'' from ITU-T~{\cite{ITUT2022Y3090,ITUT2023Y3091}}.}
    \label{fig:ITU_functional_framework}
\end{figure}

As shown in Fig.~\ref{fig:ITU_functional_framework}, ITU-T proposes a ``\textit{three-layer, three-domain, double closed-loop}'' structure for the DTN. The bottom layer is the physical network layer, which encompasses all network elements in the real system and exchanges data and control with the twin via southbound interfaces. The middle layer is the network digital twin layer, which is the core of the DTN and consists of three domains: a data domain realized by unified data repository, a model domain with basic models and functional models, and a management domain that handles topology management, model life cycle management, and security. The top layer is the network application layer, where applications such as network optimization, intent-based networking, and network visualization interact with the twin through northbound interfaces and trigger control updates back to the physical network. The ``double closed-loop'' refers to (i) an inner loop inside the twin layer, where data and models are used to emulate and optimize candidate configurations, and (ii) an outer loop spanning all three layers, where validated changes are enforced on the physical network and their impact is monitored and fed back to the twin.

\paragraph{Functional Requirements}

Efficient data collection must be supported, with the ability to capture diverse data types (e.g., equipment status, port and link statistics, key performance indicators (KPIs), user and service attributes) at appropriate frequencies and with selective sampling to avoid overwhelming network resources. The standard calls for a scalable, unified data repository capable of storing heterogeneous data  and providing services such as fast search, historical snapshots, and high-concurrency access. A further requirement is the definition of unified data models that can capture network elements, topologies, and functional behaviors (e.g., optimization, traffic management, and fault prediction), and that interface both to the data repository and to applications. Finally, the standard requires open, standardized southbound and northbound interfaces with configurable collection methods and support for high-rate information exchange and secure control signaling. 

For wireless digital twins, these functional requirements align with the need to ingest time-series RAN measurements and KPIs via standardized telemetry, maintain 3GPP-compliant configuration models for 5G Node B (gNB) and user equipment (UE), and expose twin services (e.g., what-if experiments, AI-driven optimization suggestions) to higher-level RAN automation functions.

\paragraph{Service Requirements}

Compatibility requires that DTNs support heterogeneous networks (including radio access, transport, core, and data center networks), multiple vendors, and coexistence with existing management systems. Scalability demands that DTNs cope with large-scale networks and track scaling in the physical network while maintaining performance. Reliability, security, and privacy aspects specify that (i) data and interactions between the twin and the physical network must be trustworthy, (ii) the system should withstand hardware and software failures, and (iii) customer data and operations must be protected throughout their life cycle. Additional requirements cover flexibility (on-demand data collection and model composition), visualization (multi-granular visualization of topology, traffic, and model life cycle), and synchronization (ability to track the physical network state and apply control actions within acceptable delays). 

From a wireless perspective, these requirements translate into, for example, the ability to scale a RAN digital twin from a single site to a dense urban macro deployment, to support multi-vendor gNBs and open RAN (O-RAN) components, to provide 24/7 availability for operations support, and to visualize coverage, interference, and mobility patterns at multiple spatial and temporal resolutions.

\paragraph{Capability Levels}

To facilitate the measurement of the maturity of a DTN deployment, ITU-T provides a five-level capability model for DTN systems in its Recommendation Y.3091 {\cite{ITUT2023Y3091}}.
\begin{itemize}
\item \textbf{L1 (representation level)}: the twin provides a one-way mapping from the physical network into the virtual domain. It mirrors states and behaviors with high-fidelity visualization, but all control still relies on human operators. 
\item \textbf{L2 (interaction level)} adds a control channel back to the physical network, enabling two-way interaction where operations performed in the twin can be pushed to the real network. 
\item \textbf{L3 (prediction level)} requires the twin to analyze data trends and infer unmeasured indicators or what-if scenarios, allowing it to evaluate performance and predict network evolution. 
\item \textbf{L4 (optimization level)} extends L3 with AI-based and knowledge-based optimization to support timely, intelligent decision-making and closed-loop control. 
\item \textbf{L5 (autonomy level)} represents an ideal state in which the twin and the physical network continuously perceive each other's states and autonomously reconfigure themselves while maintaining long-term consistency. 
\end{itemize}

To operationalize these levels, ITU-T defines six evaluation dimensions: (i) data service, (ii) digital twin modelling, (iii) interactive mapping, (iv) intelligence, (v) user experience, and (vi) trustworthiness. Each dimension is broken down into concrete evaluation indicators with their own internal capability levels. For example, the data service dimension includes data richness, update frequency, compatibility, data quality, data service interfaces, and processing efficiency, with levels ranging from simple equipment and status data updated daily to heterogeneous, processed data streams available in real time over standardized interfaces.

For wireless digital twins, the five-level capability model may be interpreted as follows: an offline RAN simulator fed by log data corresponds roughly to L1; a RAN digital twin that can push configuration changes into the SON/SMO framework reaches L2; adding what-if evaluation of scheduling or power-control policies yields L3; AI-driven, near real-time RAN optimization corresponds to L4; and fully self-evolving, autonomous RAN digital twins sit at the aspirational L5.

\paragraph{Integration with Management \& Orchestration in 5G}

\begin{figure*}[htbp!]
    \centering
    \includegraphics[width=6in]{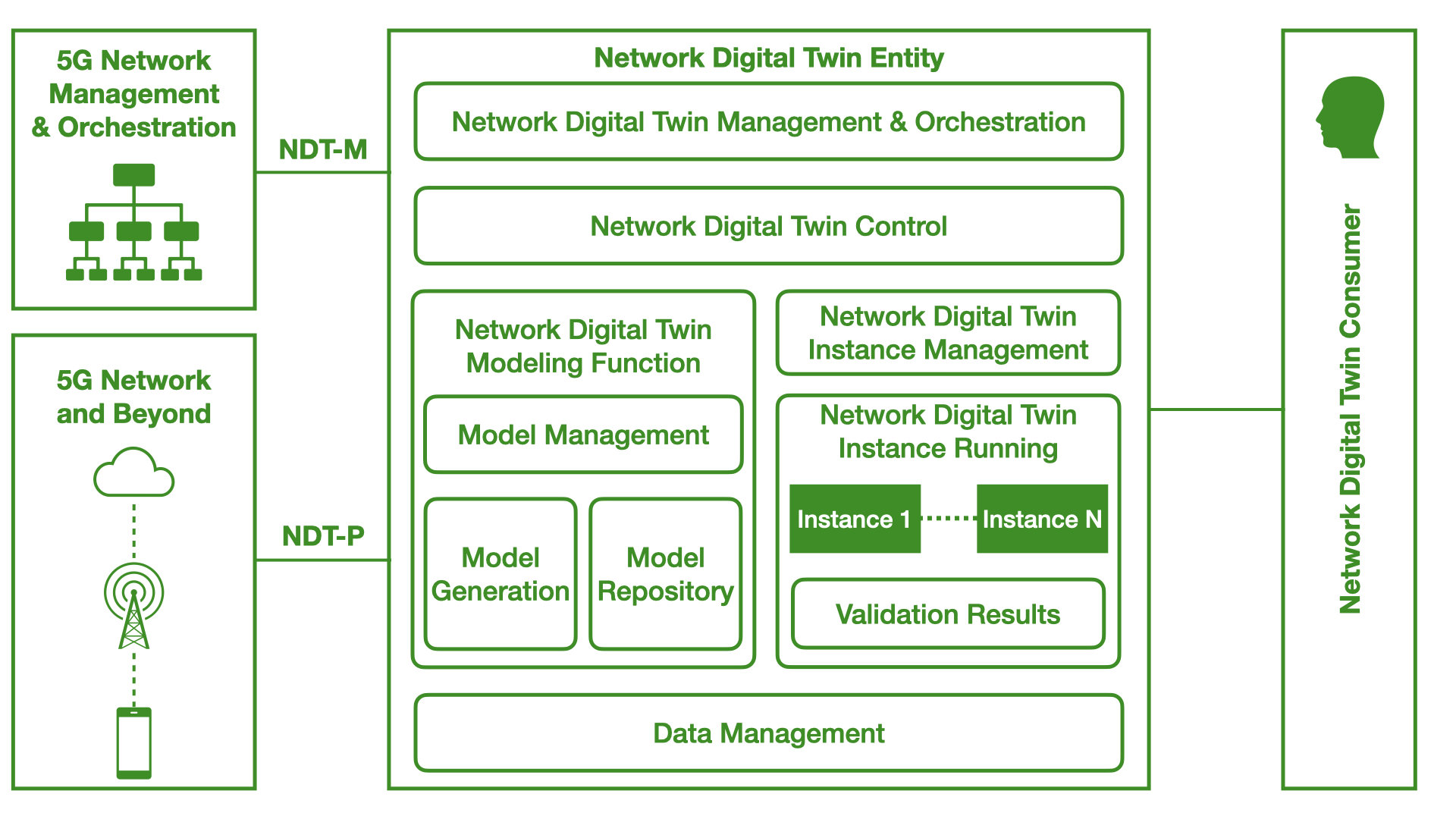}
    \caption{\small A functional framework of network digital twin based on the ITU-T recommendation Y.3092.}
    \label{fig:ndt_functional_framework}
\end{figure*}

ITU-T addresses how a network digital twin integrates with the management \& orchestration framework in 5G, by introducing two reference points, \textit{NDT-m} and \textit{NDT-p}. Figure \ref{fig:ndt_functional_framework} provides an illustration of the functional framework of network digital twin based on the ITU-T recommendation Y.3092 \cite{ITU-T-Y3092}.
The NDT-m interface carries network management data requests and responses between the 5G management \& orchestration system and the network digital twin. Specifically, performance counters, alarms, logs, and other management data are requested by the network digital twin, while the network digital twin can return recommended configurations and policies back to the 5G management \& orchestration system over the same interface. The NDT-p interface connects the network digital twin directly to the physical 5G network, carrying network status data such as node-level state and KPIs. Together, NDT-m and NDT-p formalize the southbound interface of the twin, allowing 5G RAN-specific telemetry, e.g., gNB counters, per-beam measurements, UE mobility traces, to flow into the wireless digital twin in a standardized way.

The ITU-T recommendation Y.3092 suggests that 5G networks and their management \& orchestration systems support network digital twin driven data collection from managed entities, including configuration, historical, operational, and performance data at configurable frequencies (from seconds to minutes) and with flexible methods (batched measurements vs. data streams). It further recommends that 5G networks and management \& orchestration systems support network digital twin triggered synthetic data generation and validation, so that synthetic data can be combined with real measurements. This is aligned with wireless use cases where the twin needs to generate synthetic channel or traffic traces, or diverse training datasets, which are then checked against live RAN statistics.

\subsubsection{\underline{3GPP Standardization of DTNs}}

In parallel to the ITU-T standardization of DTNs, 3GPP has been conducting standardization work on DTNs as well.

\paragraph{Management Aspect of DTNs} 

3GPP service and system aspects working group 5 (SA5) has launched a Release-19 study on the management aspects of network digital twins \cite{3GPP-TR-28915}. In this study, a network digital twin is defined as a virtual replica of a mobile network or part of one that captures its attributes, behavior, and interactions, explicitly covering both RAN and core domains. It positions network digital twins as complementary to existing network resource model (NRM) based topology and configuration models. Specifically, whereas the NRM captures what the network is, the network digital twin adds the capability to model how the network behaves under different configurations, loads and failure conditions, by means of emulation (i.e., running real network functions in a twin environment) and/or simulation (i.e., running mathematical models of those functions).

The study addresses the relationship between network digital twins and network automation functions, such as management data analytics (MDA) and SON. It treats network digital twins as behavior-modeling components that can be either internal to, or external from, these automation functions. In both cases, automation functions invoke network digital twins to evaluate the impact of candidate policies, parameter changes, or failure scenarios before they are applied to the physical network. Conceptually, this closes a gap in the current 3GPP management system: existing automation assumes that behavioral knowledge is embedded in the analytics or decision logic, whereas network digital twins provide an explicit, reusable, and shareable source of that knowledge.

3GPP investigates a set of diverse use cases relevant for wireless digital twins. 
\begin{itemize}
\item For RAN-centric optimization, a key use case is RAN energy saving policy verification using a network digital twin. Specifically, instead of iteratively tweaking energy saving policies on the real RAN, an operator requests a network digital twin instance to simulate the effect of specific energy saving policies (e.g., turning capacity-booster cells into energy saving states) on energy efficiency, UE throughput and latency, using performance measurements and configuration data defined in existing 3GPP management specifications. 
\item For resilience and reliability, the study considers use cases on signaling-storm analysis, emergency preparedness, and network failure and risk prediction. Here, network digital twins are used to replicate extreme conditions (e.g., floods of control plane messages, natural disasters, high-risk configuration changes) and to evaluate mitigation strategies in the twin before any change is made in the real network. 
\item For automation and AI, dedicated use cases cover network digital twin support to network automation and using network digital twin to generate machine learning training data. In the former case, the network digital twin acts as an off-policy evaluation engine for automation functions, and in the latter case, it acts as a synthetic data generator to augment scarce or unbalanced real-world datasets. 
\end{itemize}
Additional use cases address nested DTNs (composition of multiple twins), visualization of network topology and traffic, and configuration verification, including key event assurance for major sports events or festivals with bursty traffic.

The study also considers the life cycle management of network digital twins within the 3GPP management system. A management service (MnS) consumer (e.g., a SON function, an analytics application, or a machine learning training component) requests the creation of a network digital twin instance with given simulation or modeling requirements (e.g., scope, time horizon, data to be collected, performance requirements). The MnS producer (i.e., the entity offering the network digital twin management service) then instantiates and configures the network digital twin, synchronizes data from the managed network, executes the required emulation/simulation, and finally reports results back to the consumer.

Building on the study, 3GPP turns network digital twins  into a normative component of the 5G management system \cite{3GPP-TS-28561}. The specification defines how a network digital twin is exposed as a managed resource, how management services interact with it, and how information models represent its scope, inputs and outputs, without constraining the internal modeling technology (e.g., simulation, emulation, AI-based models). Central to the specification is the network digital twin MnS. A network digital twin MnS producer offers simulation/emulation capabilities; authorized MnS consumers (e.g., MDA function, AI inference functions, intent handlers) request that specific network scenarios be modeled and receive structured network digital twin reports as output.

\paragraph{Wireless Channel Modeling for Network Digital Twins} 

An essential component of wireless digital twins is the modeling of wireless propagation channels {\cite{lin20236g2}}. A starting point can be the long-standing 3GPP stochastic channel model in TR 38.901 {{\cite{3GPP-TR-38901}}}. This document provides a geometry-based stochastic model (GBSM) for 0.5-100 GHz, specifying large-scale parameters (e.g., path loss, shadowing, delay and angular spreads) and small-scale fading (e.g., clusters, rays, spatial consistency) for a set of canonical scenarios, including urban macro (UMa), urban micro (UMi), rural macro (RMa), indoor, etc. The model is measurement inspired, i.e., many parameters are derived from measurement and ray-tracing campaigns, but what is modeled is the stochastic recipe, not the underlying physics engine. The GBSM in TR 38.901 has become the de facto channel model for 5G link- and system-level evaluation and the basis for performance studies, but it largely abstracts away from site-specific geometry and material properties.

In Release 19, 3GPP augments this framework to support integrated sensing and communication (ISAC) \cite{lin2025bridge}. The ISAC channel modeling study item starts from the GBSM in TR 38.901 and introduces deterministic components associated with sensing targets and environment objects to properly describe mono-static and bi-static sensing channels while remaining compatible with communication-only evaluation. In practice, this means that the background propagation continues to follow the existing stochastic model, while additional deterministic target paths (parameterized, for example, by radar cross section (RCS), delay-Doppler trajectory, and angular response) are superimposed to represent reflections from drones, vehicles, humans, buildings, etc. This extended GBSM enables evaluation of ISAC KPIs such as detection probability and localization accuracy.

{
Note that TR 38.901 must not be characterized as excluding map-based ray tracing. Clause 7 specifies the widely used stochastic methodology, while Clause 8 specifies an alternative map-based hybrid model that combines deterministic ray tracing over a digitized map with stochastic components~{{\cite{3GPP-TR-38901}}}. It notes that the map-based hybrid model is not calibrated and may be used on a company basis. Thus, a ray-tracing-based methodology already exists in 3GPP, and the remaining DTN challenge lies in interoperable scene acquisition, ongoing geometry or material calibration, implementation alignment, and validation for a particular deployment. Besides, support for this hybrid model does not by itself maintain a continuously synchronized scene or certify a vendor's reconstruction accuracy.

\paragraph{ETSI closed-loop management}
ETSI GS ZSM 002 defines a reference architecture for zero-touch network and service management, while GS ZSM 009-1 defines closed-loop automation enablers~\cite{ETSI_ZSM002,ETSI_ZSM0091}. Their management domains and loop-coordination functions complement DT-based what-if evaluation. A twin can supply predictions to a management loop, where the management system still governs action execution and coordinates potentially conflicting loops. These specifications primarily provide management architecture and automation interfaces, rather than a prescribed physical twin or ray-tracing implementation.
}

\subsection{Challenges and Solutions}

The standardization and deployment of DTNs expose a set of tightly coupled technical challenges that arise from the intrinsic properties of wireless systems, the requirements of operational networks, and the constraints of large-scale management frameworks. These challenges are summarized and addressed as follows.

\subsubsection{Fidelity vs. standardization in wireless modeling}  
A primary challenge lies in balancing modeling fidelity with the need for standardization. Wireless digital twins require accurate, site-specific modeling of propagation, interference, blockage, and mobility in order to support reliable performance prediction and optimization. However, standardization bodies such as 3GPP have historically favored stochastic, scenario-based channel models due to their reproducibility, abstraction level, and implementation independence. Deterministic models, particularly ray-tracing-based approaches, offer significantly higher physical fidelity but depend heavily on detailed geometry, material properties, and numerical solvers, which vary across environments and implementations. This creates a fundamental tension between reproducible standardized models and the realism demanded by operational digital twins. The emerging solution adopted by ITU-T and 3GPP is to standardize architectural abstractions, interfaces, and data flows rather than prescribing specific channel modeling engines. This allows stochastic, deterministic, and hybrid models to coexist within the same digital twin framework, enabling fidelity where needed while preserving interoperability at the system level.

\subsubsection{Interoperability across vendors and domains}  
Operational wireless networks are inherently multi-vendor and multi-domain, encompassing radio access, core, and transport segments managed by different entities. DTNs must therefore operate across heterogeneous equipment, interfaces, and management systems. Without standardization, digital twins risk becoming vendor-specific silos. ITU-T and 3GPP address this challenge by defining open, standardized interfaces and information models while allowing internal implementations to remain vendor-specific. This approach enables interoperability at the system level while preserving innovation and flexibility at the implementation level. Importantly, it also allows digital twins to coexist with existing management systems rather than replacing them.

\subsubsection{Computational efficiency and real-time responsiveness}  
Many DTN use cases, such as RAN optimization and fault mitigation, require near real-time responsiveness. High-fidelity simulations, particularly those involving deterministic channel modeling, can be computationally intensive and incompatible with strict latency constraints. The solution adopted in practice is a combination of model hierarchy and hardware  acceleration. Simplified or surrogate models can be used for fast evaluation, while more detailed simulations are reserved for offline analysis or validation. GPU-accelerated platforms and differentiable simulation engines further reduce execution time and enable gradient-based optimization, making high-fidelity modeling more compatible with real-time operation.

\subsubsection{Alignment between standardization and practical deployment} 
Finally, there exists a structural challenge in aligning standardization efforts with rapidly evolving industrial practices. While standards favor stable, reproducible abstractions, practical digital twin deployments increasingly rely on environment-specific data, proprietary tools, and AI-driven optimization. Rather than forcing convergence at the modeling level, current standards intentionally leave room for implementation diversity. This design choice allows practical tools to evolve rapidly while remaining compatible with standardized DTN architectures, creating a flexible yet interoperable ecosystem.

Collectively, these challenges and corresponding solutions highlight a key insight: \textit{the role of standardization in DTNs is not to dictate how twins are built internally, but to define how they interact, integrate, and operate within large-scale, multi-vendor network systems}.

\subsection{Use Cases and Tools}

The ITU-T and 3GPP works discussed above stop at the architectural and management level: they define what a DTN looks like, how it is managed, and how it interfaces with network (e.g., 5G) operations, but they do not prescribe the numerical engines, ray-tracing channel models, or AI tool chains that actually power wireless digital twins. In practice, those roles are filled by concrete software platforms. In this subsection, we highlight two representative classes of tools, using NVIDIA Sionna \cite{sionna} as an example of a research-grade rapid-prototyping environment and NVIDIA Aerial Omniverse Digital Twin (AODT) \cite{aodt} as an example of a product-oriented DTN platform.

\subsubsection{Sionna}

Sionna is an open-source, graphics processing unit (GPU) accelerated library for research on communication systems. It provides link-level and system-level simulation capabilities with native machine-learning support and is implemented in Python on top of TensorFlow. The library is organized into three main modules: Sionna PHY (link-level simulation of wireless systems), Sionna SYS (system-level abstractions), and Sionna RT, a differentiable ray tracer for radio propagation modeling. This combination makes Sionna particularly well suited as a model domain engine: it can generate channels, emulate transceiver behavior, and expose differentiable end-to-end performance metrics to higher-level optimization and learning algorithms.

From a wireless digital twin perspective, Sionna serves two complementary roles. First, as a rapid-prototyping environment, it enables researchers to explore new algorithms, prototype complex communication system architectures, and integrate neural networks directly in the simulation. This aligns with 3GPP's network digital twin scope: many of the behavior models that a network digital twin would offer to automation functions can be developed and validated in Sionna before being embedded into larger twins. Second, Sionna RT addresses the ray-tracing gap in the 3GPP standardization. Sionna RT is a differentiable ray tracer for radio propagation, providing GPU-accelerated, path-based modeling of propagation in 3D scenes. Its open-source nature and documented APIs make it a good reference implementation for deterministic channel modeling: different groups can share exact scene descriptions and reproduce each other's results, or compare proprietary engines against a common baseline. Recent work has even integrated Sionna RT into full-stack simulators such as ns-3, enabling high-fidelity, geometry-based channels in protocol-level simulations. In short, Sionna RT enriches the data and model domains of the DTN layer with physics-based, site-specific channels that go beyond the 3GPP stochastic channel models.

\subsubsection{AODT}

AODT is a network digital twin platform designed to emulate RAN deployments, from a single site to city scale. AODT is part of the NVIDIA AI Aerial platform and leverages NVIDIA Omniverse and GPUs to perform near real-time ray-tracing based propagation in detailed 3D environments, combined with software-defined RAN and UE simulators. In other words, it provides much of the network digital twin layer and application layer envisaged by the ITU-T Recommendation Y.3090: realistic 3D scenes (terrain, buildings, materials), physics-based propagation, protocol-stack behavior, and APIs through which optimization and AI agents can run what-if experiments on RAN deployments.

AODT is positioned as a tool for both research and product development. It has been used by operators and vendors to create site-specific twin deployments that visualize RF propagation, optimize coverage and capacity, and validate AI-based RAN algorithms before field trials. Some vendors have integrated their own tools with AODT (e.g., using AODT as the underlying twin to generate site-specific channel for realistic RAN modeling). The ecosystem integration echoes the DTN collaboration concepts in 3GPP TS 28.561, where multiple DTN functions can interoperate via proper interfaces. AODT's scene importer and use of open scene formats resemble the data domain and model domain abstractions: geometry and material data form part of the DTN's data repository, while propagation and RAN models form the basic and functional models. In particular, AODT may be wrapped by the network digital twin management services and connected to 5G management \& orchestration via the NDT-m/NDT-p interfaces, turning a product tool into a standards-aligned network digital twin component.

\subsection{Summary of Lessons Learned}

In summary, several key lessons emerge from the study of standardization and enabling tools for wireless digital twins. First, standardization is essential for bridging conceptual digital twin visions and deployable wireless systems, as it enables interoperability, scalability, and closed-loop control. Second, ITU-T and 3GPP provide complementary frameworks that define architectures, capability levels, and management interfaces while deliberately avoiding constraints on internal modeling implementations. Third, a persistent gap remains between standardized stochastic channel models and the site-specific fidelity required by practical digital twin applications. Fourth, deterministic and hybrid modeling approaches, particularly ray tracing, are indispensable for high-fidelity digital twins but must be integrated through standardized abstractions rather than mandated engines. Finally, emerging tools demonstrate that physics-based modeling, AI-native optimization, and standardized management can converge into interoperable, multi-vendor wireless digital twin ecosystems that are both operationally relevant and future-proof.

\section{Conclusion}
DTNs are emerging as a foundational paradigm for future wireless systems, enabling the convergence of communication, sensing, computation, and intelligent control within a unified cyber-physical framework. Unlike traditional DTs that primarily serve monitoring and simulation purposes, next-generation DTNs are evolving toward application-aware, semantic-driven, and closed-loop architectures capable of continuously interacting with and optimizing their physical counterparts. In this tutorial, we reviewed the fundamental principles of DTN creation, resource optimization, physical-layer design, and feedback-driven network management, while highlighting the growing role of AI, semantic communications, integrated sensing and communication, and distributed intelligence in shaping DTN evolution. We also discussed standardization efforts, practical tools, and key deployment challenges that must be addressed to realize large-scale DTN ecosystems. Looking forward, advancing DTNs will require new theories and systems that jointly address cyber-physical synchronization, semantic abstraction, trustworthy AI, distributed orchestration, and real-time closed-loop adaptation. We envision that DTNs will play a native role in enabling future 6G and beyond networks, transforming wireless infrastructures from passive communication platforms into intelligent and autonomous systems that continuously sense, learn, reason, and optimize across diverse application domains.

\bibliographystyle{IEEEtran}
\begingroup\color{black}\bibliography{refs}\endgroup

\end{document}